# Exchanging Ohmic Losses in Metamaterial Absorbers with Useful Optical Absorption for Photovoltaics

**Ankit Vora[1], Jephias Gwamuri[2], Nezih Pala[3], Anand Kulkarni[1], Joshua M. Pearce[1,2], and Durdu Ö. Güney[1,*]**

[1]Department of Electrical and Computer Engineering, Michigan Technological University, Houghton, MI 49931, USA
[2]Department of Materials Science and Engineering, Michigan Technological University, Houghton, MI 49931, USA
[3]Department of Electrical and Computer Engineering, Florida International University, Miami, FL 33174 USA
[*]Corresponding Author: dguney@mtu.edu

**Abstract** Using metamaterial absorbers, we have shown that metallic layers in the absorbers do not necessarily constitute undesired resistive heating problem for photovoltaics. Tailoring the geometric skin depth of metals and employing the natural bulk absorbance characteristics of the semiconductors in those absorbers can enable the exchange of undesired resistive losses with the useful optical absorbance in the active semiconductors. Thus, Ohmic loss dominated metamaterial absorbers can be converted into photovoltaic near-perfect absorbers with the advantage of harvesting the full potential of light management offered by the metamaterial absorbers. Based on experimental permittivity data for indium gallium nitride, we have shown that between 75% − 95% absorbance can be achieved in the semiconductor layers of the converted metamaterial absorbers. Besides other metamaterial and plasmonic devices, our results may also apply to photodetectors and other metal or semiconductor based optical devices where resistive losses and power consumption are important pertaining to the device performance.

Improvements in optical enhancement for solar photovoltaic (PV) devices using conventional natural materials has begun to show diminishing returns, yet metamaterials, which are rationally designed geometries of optical materials that can be tuned to respond to any region of the electromagnetic spectrum offer an opportunity to continue to improve solar device performance[1]. Historically, periodic structures with sub-wavelength features are used to construct a metamaterial for a given application. Designing and constructing a material in this way a metamaterial possesses optical properties that are not observed in their constituent materials enabling the properties to be determined from structure rather than merely composition. The advent of metamaterials has provided PV device designers among others with unprecedented flexibility in manipulating light and producing new functionalities[2]. Metamaterials have already been proposed for super lenses that allow sub-wavelength resolution beyond the diffraction limit, and electromagnetic cloaks, which promise real physical invisibility[3].

Another approach to optical enhancement of PV is to consider plasmonics, which is a rapidly growing field for the application of surface plasmons to device performance improvement. Surface-plasmon-based devices already proposed include not only PV but also light-emitting devices, data storage, biosensors, nano-imaging, waveguides, perfect absorbers, and others[4-8].

Surface plasmons are collective oscillations of surface electrons whereas surface plasmon polariton (SPP) describes a coupled state between a surface plasmon and a photon[9]. On the other hand, localized surface plasmons (LSP) are confined to bounded geometries such as metallic nanoparticles or nanostrips of various topologies[10-12]. The fluctuations of surface charge results in highly localized and significantly enhanced electromagnetic fields in the vicinity of metallic surfaces.

Typically, surface plasmon resonances exhibit a strong relationship to the size, shape and the dielectric properties of the surrounding medium. The resonances of noble metals are mostly in the visible or infrared region of the electromagnetic spectrum, which is the range of interest for PV applications[13].

For the application, we are considering here photons arriving at the metal surface produce surface waves in the form of SPP along the metal-semiconductor interface of the top layer of a PV device. This occurs when the photons interact with the collective oscillations of free electrons in the metal of the absorber.

The metallic nanostructures have the ability to maintain SPPs, which provides electromagnetic field confinement and enhancement. This phenomenon has already attracted substantial attention in the PV scientific community, but has yet to be commercialized and scaled.

Perhaps even more promising it is possible to construct plasmonic metamaterial "perfect absorbers" with such surface waves. Metamaterial absorbers are extremely flexible and have already been designed with broadband[14-19], polarization-independent[17-21], and wide-angle[14, 15, 17, 19, 20, 22] optical absorption for a host of applications. Both theoretical work through device simulations[15, 18, 22] and physical fabrication[14, 16, 17, 19-21, 23] of ultrathin metamaterial absorbers with these features have been realized. The theoretical limit of nearly 100% absorbance at the tunable resonant wavelength can be obtained with proposed designs of ultrathin, wide-angle perfect absorber structures for infrared light[22]. Very recently, a study showed an average measured absorption of 71% from 400 – 700nm, which represents a majority of the energy in the solar spectrum, with an ultrathin (260nm), broadband and polarization independent plasmonic super absorber[19]. In addition, the simulations of Aydin et al., indicated that the absorption levels could be increased to 85%[19].

All of these features are critical to maximize the efficiency of PV devices, yet are generally lacking in schemes to optically enhance solar cell design. Even small improvements in optical enhancement in specific cases can have a large effect. For example, improving (enlarging) the acceptance angle for high absorption for a PV device may eliminate the need for mechanical tracking of a PV system, which can have a substantial impact on both the first cost as well as the operating and maintenance costs.

Complex perfect absorbers have more degrees of freedom in terms of impedance-match, polarization-independence, and wide angular reception, which make them superior systems for managing light than simple plasmonic nanoparticle approaches. Thus perfect absorbers are better candidates for enhancing solar PV conversion efficiency.

Despite this promise, perfect absorbers have not been studied systematically in the literature for solar photovoltaic applications[1]. In addition, in the few cases where the potential was probed the device design limited the practicality of the approach. For example, Aydin, *et al*.'s "super absorber"[19], uses a lossless spacer where almost all the power covering the solar spectrum is absorbed by the metal. This results in significant Ohmic loss via heating, which loses both the power of the photons but also decreases the efficiency of the next electrical conversion because of resultant higher cell operating temperature. Instead, the approach taken in this paper is to change the absorption localization from metal to the semiconductor layer by replacing the dielectric spacer with a "semiconductor spacer" in the perfect (or "super") absorbers, which can then be used to enhance thin-film PV device performance.

The geometric skin depth (i.e., depth at which the current density is reduced to $1/e$ of its value at the surface of the conducting material) in a structured conductor such as metallic nanostructures (e.g., split-ring resonators, strips, etc.) is much larger than in the bulk form of the same conductor[24]. However, in the limit the nanostructure approaches bulk conductor, the geometric skin depth is significantly reduced and approaches its bulk value. More importantly, this skin depth reduction can be exploited as an important resource to minimize Ohmic losses in metallic nanostructures despite higher metallic filling ratio. We will call this technique the "Bulk Skin Depth Technique (BSDT)." Previously, we applied this technique to reduce Ohmic losses in metamaterials[24] and to design three-dimensionally isotropic negative index metamaterial[25] (see Fig. S1, Supplementary information). Now, in this paper we apply the BSDT to perfect absorbers to turn them into functional solar cell devices and use the indium gallium nitride ($In_xGa_{1-x}N$) PV device as an example.

If we reduce the geometric skin depth in the metallic portions of the metamaterial absorbers (either perfect or near-perfect), we can push the resonant currents induced by solar radiation closer to the semiconductor layers in PV devices. This brings the following important advantages: (1) Optical absorption in the metallic layers (i.e., Ohmic loss) is shifted into semiconductors, which increases useful optical absorption for solar energy conversion. (2) Joule heating is minimized. (3) Because the current is tightly confined to metallic surfaces facing the semiconductor layers, additional interconnects or contacts can be made possible[25] without destroying or short-circuiting the absorber (see Fig. S1, Supplementary information).

In order for a plasmonic PV cell to reach the full potential of the technology the designed cell must maximize the absorption within the semiconductor region. Simultaneously, absorption must be minimized everywhere else including the metallic regions over the cell. This balancing act of absorption must be accomplished over the entire solar spectrum (normally AM1.5, which is ASTM G-173) and ideally over all incidence angles. Previous work has shown that a BSDT[24] can be employed to minimize Ohmic losses in metamaterial absorbers by guiding optical absorption to the semiconductor layers and away from the metal. In order for this technique to function, it first requires identification of resonant modes of the metamaterial absorber that contributes to optical absorption. Then the nanostructure of the metal must be geometrically customized as determined by the underlying resonant currents.

The BSDT approach can be applied to different geometries (e.g. a square-grid structure made up of cross structures formed by perpendicularly placed metallic strips) or any other polarization-independent metamaterial absorber design found in the literature[15, 17, 23, 26, 27]. In addition this approach will work for both "super absorber"[19] as well as metamaterial absorbers with broadband response employing the concept of nanoantenna[14, 15].

**Results**

**Applying the BSDT to a metamaterial perfect absorber and exchanging Ohmic losses.** Below we show how to apply the BSDT to a simple perfect absorber to illustrate the concept. Application to other perfect absorbers is similar. In Fig. S1 (see Supplementary information) we provide a different (earlier) example where the BSDT is applied to split-ring-resonators with resonant loop currents[24]. Figs. 1(a)-(c) show a two-dimensional cross section of the simple metamaterial absorber, induced resonant currents, and power loss density, respectively. The absorber consists of gold (red) layers and a lossless spacer (blue).

In most metallic nanostructures operating around optical frequencies the geometric skin depth is larger than both the bulk skin depth and the dimensions of the cross-sectional area of the metal through which induced currents flow. This usually results in larger Ohmic losses at optical frequencies compared to lower frequencies due to deeper penetration of the photons relative to structure dimensions[24]. Therefore, it is desired to reduce the geometric skin depth at optical frequencies to reduce the Ohmic losses. The BSDT offers an effective strategy for the geometric tailoring of the skin depth. In order to apply the BSDT we first need to identify the underlying resonant mode of the nanostructure and then modify the geometry such that, for induced resonant currents, the metallic surfaces of the nanostructure behave effectively similar to the surfaces of semi-infinite bulk metals with bulk skin depth. Below, we explain how the BSDT is applied to the metamaterial absorber sketched in Fig. 1(a). Fig. 1(b) shows the current density distribution for the identified plasmonic resonant magnetic dipole mode responsible for perfect absorbance. The dipole mode has antisymmetric current flow at the top metal strip and bottom ground plate indicated by black arrows. The current density has a maximum at the metal-spacer interface, where the surface plasmons reside, and decays exponentially inside the metal with the increasing distance from the interface. Fig. 1(f) shows that when $t_s = 20$nm the geometric skin depth for the strip is 38nm which is larger than both bulk skin depth (i.e., 23nm) and the cross-section of the strip through which the current flows. To reduce the skin depth we need to tailor the geometry such that the finite metal surfaces, where the resonant currents are bounded to, approach to bulk metal surfaces so that the geometric skin depth approaches bulk skin depth (or even below). For the given structure in Fig. 1(a), without an important shift in the resonance frequency, this can only be achieved by increasing $t_s$ and $t_g$. Although, at first, increasing the $w$ might also be seen accommodating for the "bulk limit," we cannot change $w$ since it is one of the critical parameters to tune the resonance frequency of the magnetic dipole as we will explain later when we discuss the scalability of the perfect absorber toward the red edge of the visible spectrum.

Fig. 1(d) shows near 100% total absorbance achieved by the simple perfect absorber using the same design parameters as Figs. 1(a)-(c) except the ground plate thickness $t_g = 60$nm to prevent transmittance. When the spacer is assumed to be a lossless dielectric, the total absorbed optical power is shared only between the top and bottom gold layers. Using BSDT, we reduce the total absorbance in the metals to about 70% as shown in Fig. 1(e). Here, BSDT is applied by making metallic layers thicker along the direction perpendicular to the flow of conduction currents; because this allows the metallic surfaces behave like semi-infinite bulk metal surfaces and reduce the geometric skin depth for the induced resonant currents, hence reduce the Ohmic losses similar to Fig. S1 (see Supplementary information).

Orange line (diamond symbols) in Fig. 1(f) shows the geometric skin depth for the strip approaches to bulk skin depth as the thickness of the strip increases. At the same time, the peak absorbance in the strip drops from about 70% at $t_s = 20$nm [see Fig. 1(d)] to about 35% at $t_s = 85$nm [see Fig. 1(e)]. The ground plate thickness is fixed at $t_g = 75$nm during the process except at the last data point where $t_g = 85$nm. This does not affect the skin depth trends for the strip or the ground plate as they have already converged to the bulk skin depth. The blue horizontal line (square symbols) corresponds to the geometric skin depth for the ground plate which is almost constant and about the same as bulk skin depth due to sufficiently large thickness. We should note that our observation in Fig. 1(f) is consistent with Fig. 7 in Ref. **24**.

The most of the unabsorbed power in Fig. 1(e) is reflected due to impedance mismatch and the localization of some absorbed power is internally changed from the strip to the ground plate since BSDT is mainly applied to the strip here. Note that the strip thickness increases 65nm during the BSDT process while the ground plate thickness increases only 10nm. In general, if we apply the BSDT to only a subset of metallic components, internal change of absorption localization from those metallic components to other metallic components occur as we observe here. This internal redistribution of power dissipation occurs at a lesser degree if BSDT leads to impedance mismatch.

We can take advantage of this impedance mismatch in the next stage by replacing the lossless spacer with a lossy spacer (such as a semiconductor). For example, if we introduce a non-zero imaginary part to the permittivity of the original lossless spacer [see Fig. 1(g)], the reflected power can be acquired back into the lossy spacer, hence not only the Ohmic losses are reduced but also the optical absorption is restored back to nearly 100%, maximized in the lossy spacer [compare with Figs. 1(d)-(e)]. The lossy spacer can improve the impedance mismatch which was created by BSDT. If the imaginary part of the electric permittivity of the lossy spacer is sufficiently large, almost all the power originally absorbed by the metal can be shifted to the spacer as discussed in our paper.

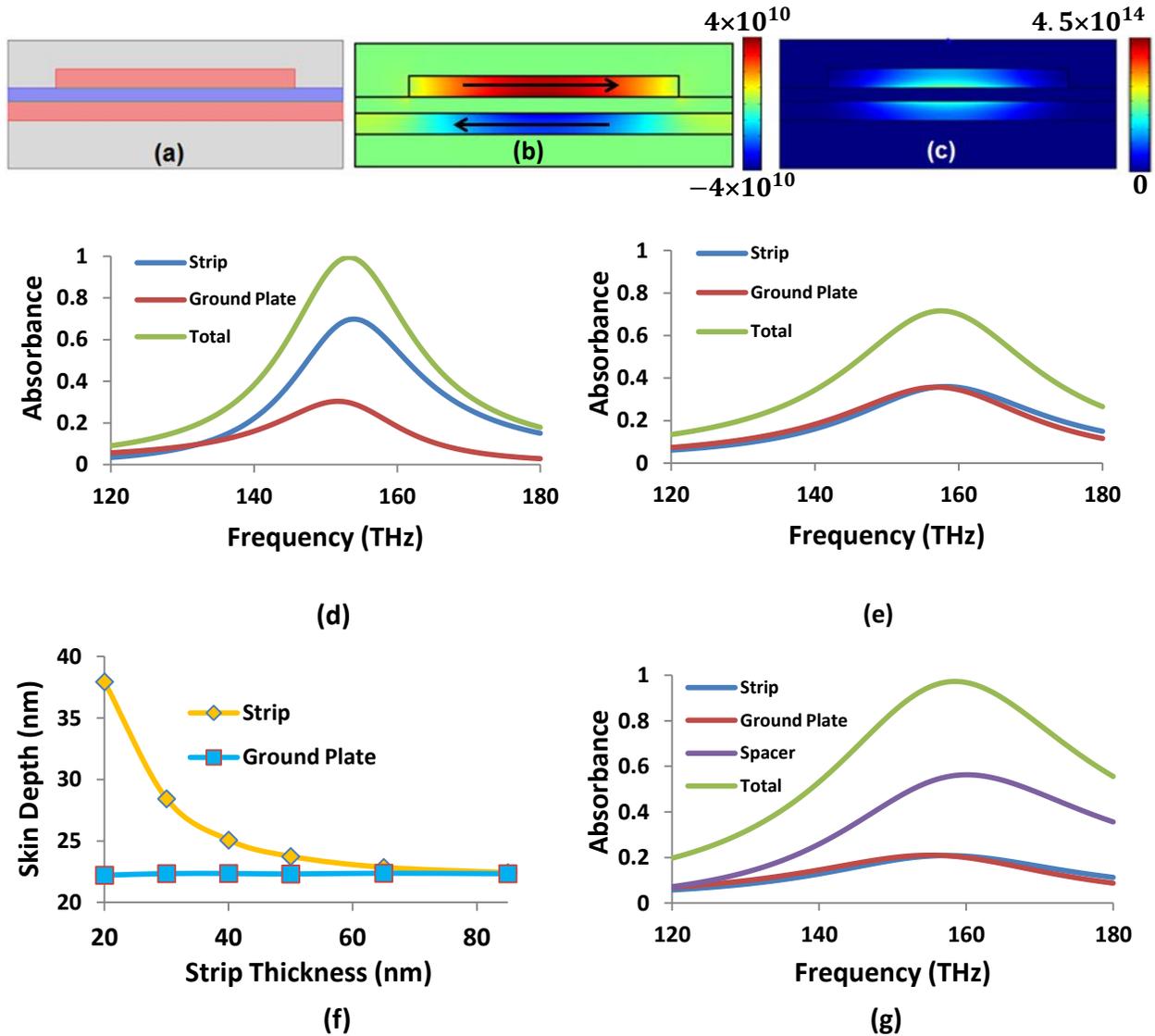

**Figure 1** (**a**) Two-dimensional cross-section of a unit cell of a simple absorber consisting of a lossless spacer (shown blue) sandwiched between a strip and a ground plate both made from gold (shown red). The size of the unit cell along the direction of periodicity is $p = 350$nm. The width of the strip is $w = 250$nm. The strip and the ground plate have the same thickness of $t_s = t_g = 20$nm. The spacer thickness is $d = 14$nm. The electric permittivity of the spacer is $\varepsilon = 2.25$. (**b**) Resultant magnetic dipole oscillating at 150THz. The antiparallel current density (A/m$^2$) indicated by arrows give rise to optical absorption in metallic regions as evidenced by (**c**) the power loss density (W/m$^3$). (**d**) Absorbance at different regions of the perfect absorber with the same design parameters as (a)-(c) except $t_g = 60$nm to prevent transmittance. Total absorbance is almost 100% shared between the strip (~70%) and the ground plate (~30%), since lossless spacer is assumed. (**e**) Using BSDT total absorbance in the metal is reduced to ~70% shared equally between the strip and the ground plate. BSDT is applied by increasing metallic thicknesses to $t_s = t_g = 85$nm. (**f**) Geometric skin depths for the strip and the ground plate for different strip thicknesses. The ground plate has a fixed thickness of $t_g = 75$nm except the last data point where $t_g = 85$nm. Geometric skin depth converges to bulk skin depth with increasing metallic thicknesses. (**g**) Introducing a non-zero imaginary part to the permittivity of the original lossless spacer in (d)-(e) (e.g., $\varepsilon = 2.25 + 0.2j$) leads to the change of absorption localization from metals to the spacer. This lossy spacer can be replaced with a semiconductor material for solar cell applications to exchange metallic Ohmic losses with useful optical absorption for carrier generation in the semiconductor regions.

**Metamaterial absorbers with In$_x$Ga$_{1-x}$N spacers.** To explore the applicability of the technique to the semiconductors used in PV technology, we chose In$_x$Ga$_{1-x}$N as the spacer because 1) it has a potential band gap engineering due to variable band gap from 0.7 to 3.4 eV created by changing the indium concentration that covers nearly the entire solar spectrum, 2) it has a high absorption coefficient that would make it possible to utilize thin absorber layers most applicable to this method of optical enhancement, 3) recent microstructural engineering has been demonstrated with nanocolumns that offer optical enhancement, and 4) this provides the potential for several known photovoltaic device configurations and multi-junctions with theoretical efficiencies over 50% [28-30].

The experimental complex electric permittivity ($\varepsilon$) data for In$_{0.54}$Ga$_{0.46}$N[31] is given in Fig. S2 (see Supplementary information). We applied the BSDT at five specially selected frequencies. These frequencies along with the corresponding $\varepsilon$ for In$_{0.54}$Ga$_{0.46}$N are listed in Table 1. If it is desired, it should be possible to apply the technique to any other intermediate frequencies between these five representative frequencies. Because these frequencies are selected carefully by considering the minima, maxima and profile of the complex permittivity data as described below, the absorbance results for the intermediate frequencies should be interpolations of the results obtained from the selected frequencies.

| Data Point | Frequency (THz) | $\Re(\varepsilon)$ | $\Im(\varepsilon)$ | $p$ (nm) | $w$ (nm) | $t_s$ (nm) | $t_g$ (nm) | $d$ (nm) |
|---|---|---|---|---|---|---|---|---|
| Point 1 | 145 | 0.86 | 1.23 | 400 | 315 | 15 | 60 | 14 |
| Point 2 | 387 | 4.59 | 0.45 | 150 | 55 | 10 | 60 | 12 |
| Point 3 | 695 | 5.67 | 1.61 | 100 | 40 | 20 | 60 | 31 |
| Point 4 | 1000 | 5.06 | 2.74 | 310 | 47 | 55 | 60 | 20 |
| Point 5 | 1090 | 5.47 | 2.60 | 282 | 57 | 70 | 70 | 18 |

**Table 1** Real and imaginary parts of the complex permittivity for the In$_{0.54}$Ga$_{0.46}$N data in Fig. S2 (see Supplementary information) at five selected frequency points and geometric parameters for the five perfect (or near-perfect) absorbers operating near those five respective selected frequencies.

The complex permittivity data in Fig. S2 (see Supplementary information) starts from 145THz, at which the permittivity is $\varepsilon_1 = 0.86 + 1.23j$. This is selected as the first frequency point. Above this point, the $\Im(\varepsilon)$ monotonically decreases and reaches global minimum at 387THz while the $\Re(\varepsilon)$ manifests a monotonic increase. Thus, the permittivity at this second selected frequency point becomes $\varepsilon_2 = 4.59 + 0.45j$. The third selected frequency point is reached at 695THz where the permittivity is $\varepsilon_3 = 5.67 + 1.61j$. This is the frequency point where $\Re(\varepsilon)$ reaches a global maximum and the $\Im(\varepsilon)$ still tends to increase. At the fourth selected frequency point, 1000THz, the $\Re(\varepsilon)$ reaches a minimum while the $\Im(\varepsilon)$ reaches global maximum. The permittivity at this point becomes $\varepsilon_4 = 5.06 + 2.74j$. The experimental data ends at 1090THz where the permittivity becomes $\varepsilon_5 = 5.47 + 2.60j$. This is selected as the fifth special frequency point.

Table 1 also lists geometric parameters for five metamaterial absorbers tuned to operate around the five respective selected frequencies (referred as Points 1-5) listed in the table. The geometry for the assumed metamaterial absorbers and the definitions of the geometric parameters are described in Figs. 1(a)-(c). The spacers used in the absorbers are initially chosen to be lossless and their permittivity is equal to the $\Re(\varepsilon)$ at the respective selected frequency. Below we show that these metamaterial absorbers can be turned into near-perfect "solar absorbers" where most of the metallic absorption is shifted to semiconductor layers (i.e., $In_xGa_{1-x}N$) by using BSDT followed by replacing the lossless spacers with the semiconductor spacers.

**Exchanging Ohmic losses at the infrared spectrum.** First, we apply the BSDT at around Point 1 to shift the optical absorption in the metallic regions to the semiconductor layer. We start from a perfect absorber which has a peak absorbance at about 180THz [see Fig. 2(a)], relatively close to Point 1. The thickness of the ground plate was chosen as 60nm to prevent transmittance. The other geometric parameters for this perfect absorber are given in Table 1. Figs. 2(b) and (c) show the magnetic field and current density distribution, respectively, for the resonant magnetic dipole mode leading to the perfect absorption. Relatively simple field profile of the resonant mode renders the application of BSDT fairly straightforward. Because the surface plasmons are mainly confined to metal-spacer interfaces, we can push the surface plasmons further to the interface by increasing the thicknesses of both the strip and the ground plate. This will reduce the skin depth and prevent further penetration of the electromagnetic fields into the metallic regions. When the strip thickness reaches from its original value of 15nm to the same thickness as the ground plate, the photons cannot penetrate deep into the strip anymore and hence the peak absorbance in the strip reduces from about 75% to 30% [see Fig. 2(d)]. The photons which are not absorbed anymore by the strip are either reflected or added to the absorbance in the ground plate. The overall peak absorbance of about 60% is shared almost equally by the strip and the ground plate at the same thicknesses of 60nm. The overall final peak absorbance can be reduced further down to about 48%, again shared almost equally by the strip and the ground plate when they both reach 245nm thickness [see Fig. 2(e)]. The unabsorbed photons in the process are reflected.

Having minimized the metallic absorbance in the perfect absorber by BSDT, in the next stage we gradually restore the $\Im(\varepsilon)$ of $In_{0.54}Ga_{0.46}N$ in Fig. 3. This leads to the total absorbance in the metallic regions being further minimized to below 15% at the first targeted frequency 145THz. The photons expelled from the metallic regions as well as photons reflected originally due to impedance mismatch with the increasing metal thicknesses during the application of BSDT are now shifted to and absorbed in the semiconductor layer (see Fig. 3). The total peak absorbance in the semiconductor layer at the target frequency exceeds 85% and the overall perfect absorption capability of the absorber is restored when the $\Im(\varepsilon)$ takes its true experimental value of 1.23. In contrast, the maximum total absorbance in plain $In_{0.54}Ga_{0.46}N$ layer for the same thickness (i.e., $d = 14$nm) is only about 6% based on available experimental complex permittivity data in Fig. S2 (see Supplementary Information). Thus, we convert the Ohmic loss dominated perfect absorber to "photovoltaic near-perfect absorber." In this stage, the resonant surface plasmon mode displayed in Figs. 2(b) and (c) again plays an important role on changing absorption localization from the metallic regions to the semiconductor layer. Since the surface plasmons reside at the metal-semiconductor interface, the accompanying photons can be easily absorbed near the interface by the semiconductor layer. This process of shifting absorption localization to the semiconductor layers can be achieved over the entire infrared spectrum under the same strategy above using the same magnetic dipole mode.

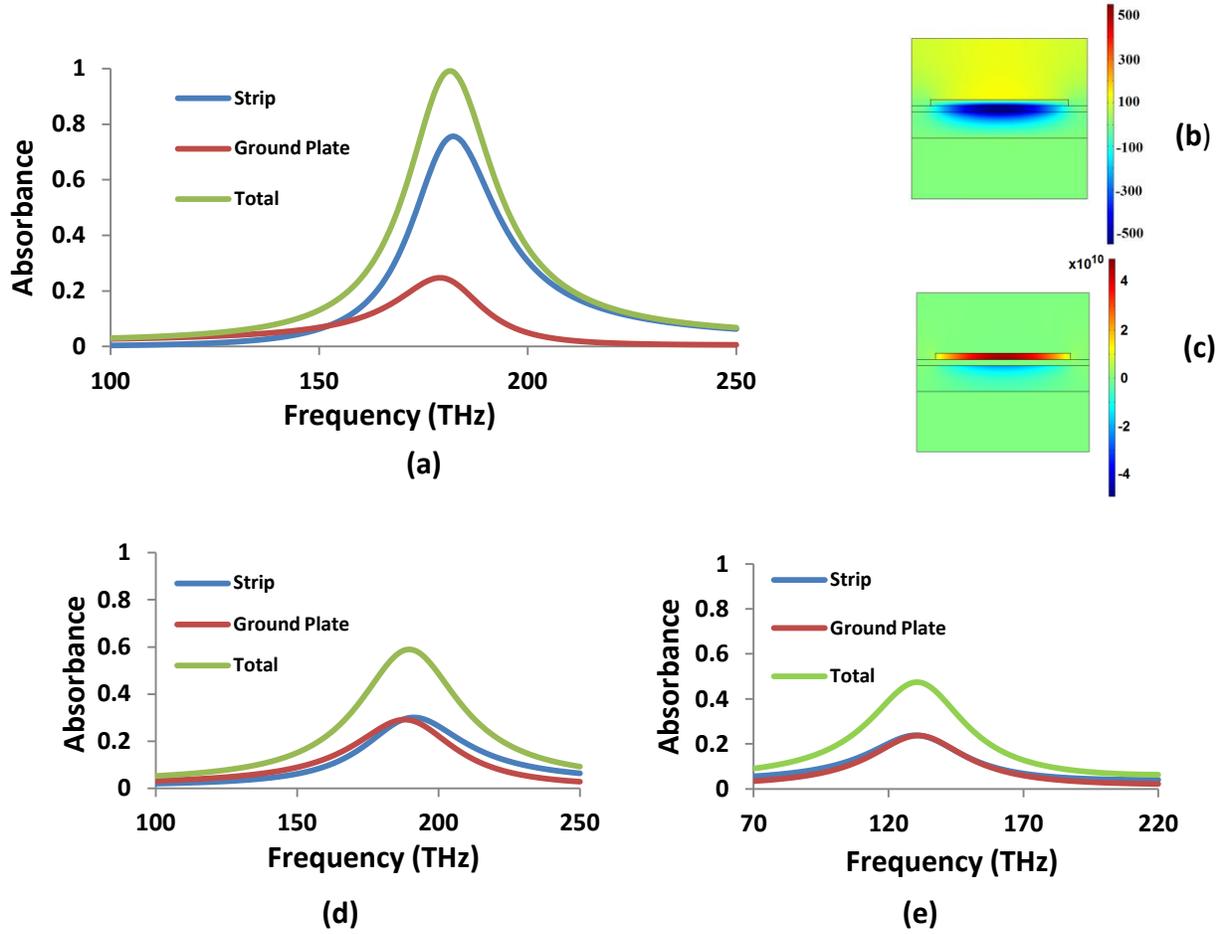

**Figure 2** (**a**) Absorbance in different parts of the perfect absorber operating in the neighborhood of Point 1 in Table 1. (**b**) Magnetic field (A/m) component perpendicular to the cross section (i.e., defined as $H_z$) and (**c**) distribution of the current density (A/m$^2$) parallel to the metal-spacer interface (i.e., defined as $J_x$) at the peak absorbance. Red and blue colors correspond to the current flowing to the right and left, respectively. (**d**) Absorbance in different parts of the absorber when the strip thickness increases to $t_s = 60$nm and (**e**) when both the strip and the ground plate have $t_s = t_g = 245$nm thickness. All the other parameters are kept fixed and the same as in Table 1.

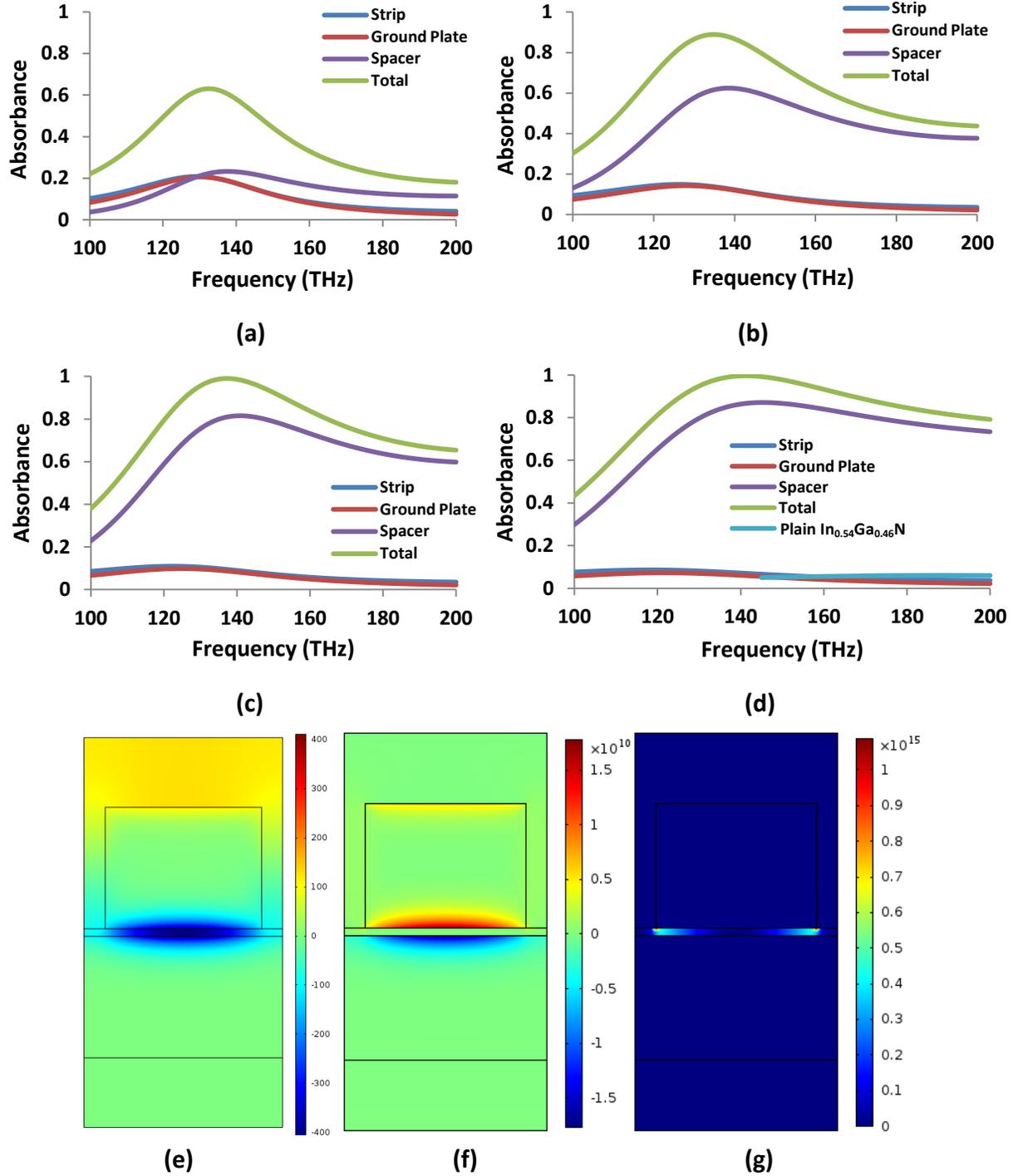

**Figure 3** Absorbance in different parts of the absorber achieved by the BSDT process described in Fig. 2 followed by the gradual incorporation of the true complex permittivity such that (**a**) $\Im(\varepsilon) = 0.1$, (**b**) $\Im(\varepsilon) = 0.4$, (**c**) $\Im(\varepsilon) = 0.8$, and (**d**) $\Im(\varepsilon) = \Im(\varepsilon_1) = 1.23$. The total absorbance in plain $In_{0.54}Ga_{0.46}N$ layer of the same thickness (i.e., $d = 14$nm) is also shown for comparison using the experimental complex permittivity data in Fig. S2 (see Supplementary information) (**e**) $H_z$ (A/m), (**f**) $J_x$ (A/m$^2$), and (**g**) the power loss density (W/m$^3$) distribution for the near-perfect absorber in (d) for the first selected frequency of 145THz in the $In_{0.54}Ga_{0.46}N$ data (i.e., Point 1). Significant portion of the power (i.e., over 85%) is absorbed inside the $In_{0.54}Ga_{0.46}N$ layer.

Below, we study the role of BSDT in more detail to better understand the transition from perfect metallic absorbance in Fig. 2(a) to near-perfect semiconductor absorbance in Fig. 3(d). Fig. 4(a) shows absorbance in different parts of the absorber and normalized skin depth (i.e., normalized with respect to its maximum value of 53nm) versus $t_s$ in the same graph. The skin depth is maximized at small $t_s$ values and decays exponentially with increasing $t_s$ and finally converges to bulk skin depth values when the $t_s$ becomes sufficiently large. The change in the skin depth has important impact on the surface impedance[15, 32] of the metamaterial absorber. Below, we show that tailoring the geometric skin depth can be exploited as an important and convenient resource to control the surface impedance and absorbance in different parts of the absorbers.

As shown in Fig. 2(a) the structure behaves as an impedance matched perfect absorber with the lossless spacer when $t_s = 15$nm. Because, when $t_s = 15$nm, the skin depth is 53nm (i.e., larger than the $t_s$). This allows photons to easily penetrate through metal and be completely absorbed as shown in Fig. 2(a). However, when the $t_s$ becomes larger, the metal strip approximates bulk metal and thus the skin depth for the resonant magnetic dipole mode decreases and approaches to bulk skin depth as shown in Fig. 4(a). Since the skin depth is reduced, photons cannot penetrate deeper into the metal. For example, when $t_s = 35$nm, the skin depth is reduced to 32nm (i.e., less than the $t_s$). This not only reduces the absorbance but also increases the impedance mismatch. Increasing the $t_s$ further above 60nm results in converging skin depth. Since the skin depth is converged, both the absorbance and impedance mismatch also tend to converge as can be seen from absorbance plots in Fig. 2. In the following we explain that the impedance mismatch achieved by tailoring the skin depth can be used to control where the absorbance is localized.

The absorbance curves in Fig. 4(a) are obtained by replacing the lossless spacer in Fig. 2 with the lossy spacer such that $\Re(\varepsilon) = \Re(\varepsilon_1) = 0.86$ and $\Im(\varepsilon)$ is assumed to have four different values which are 0.1, 0.3, 0.6, and $\Im(\varepsilon_1) = 1.23$, to illustrate the interplay between the skin depth, imaginary part of the permittivity of the lossy spacer, and impedance matching. At $t_s = 15$nm we observe that the peak absorbance in the lossy spacer is decreased with increasing $\Im(\varepsilon)$, because the absorber [see Fig. 2(a)] at $t_s = 15$nm with lossless spacer was an impedance matched perfect absorber. When we increase the imaginary part of the lossy spacer, the structure becomes increasingly impedance mismatched. Therefore, the absorbance in both lossy spacer and metals decreases with increasing $\Im(\varepsilon)$. Fig. 4(b) shows surface impedance for $\Im(\varepsilon_1) = 1.23$ and normalized skin depth versus $t_s$ in the same graph. Notice that in the region where the skin depth is high, the structure is impedance mismatched and starts to become impedance matched as the skin depth converges. (See Fig. S3, Supplementary information, to analyze the impedance matching for different $\Im(\varepsilon)$ values based on total absorbance plots.)

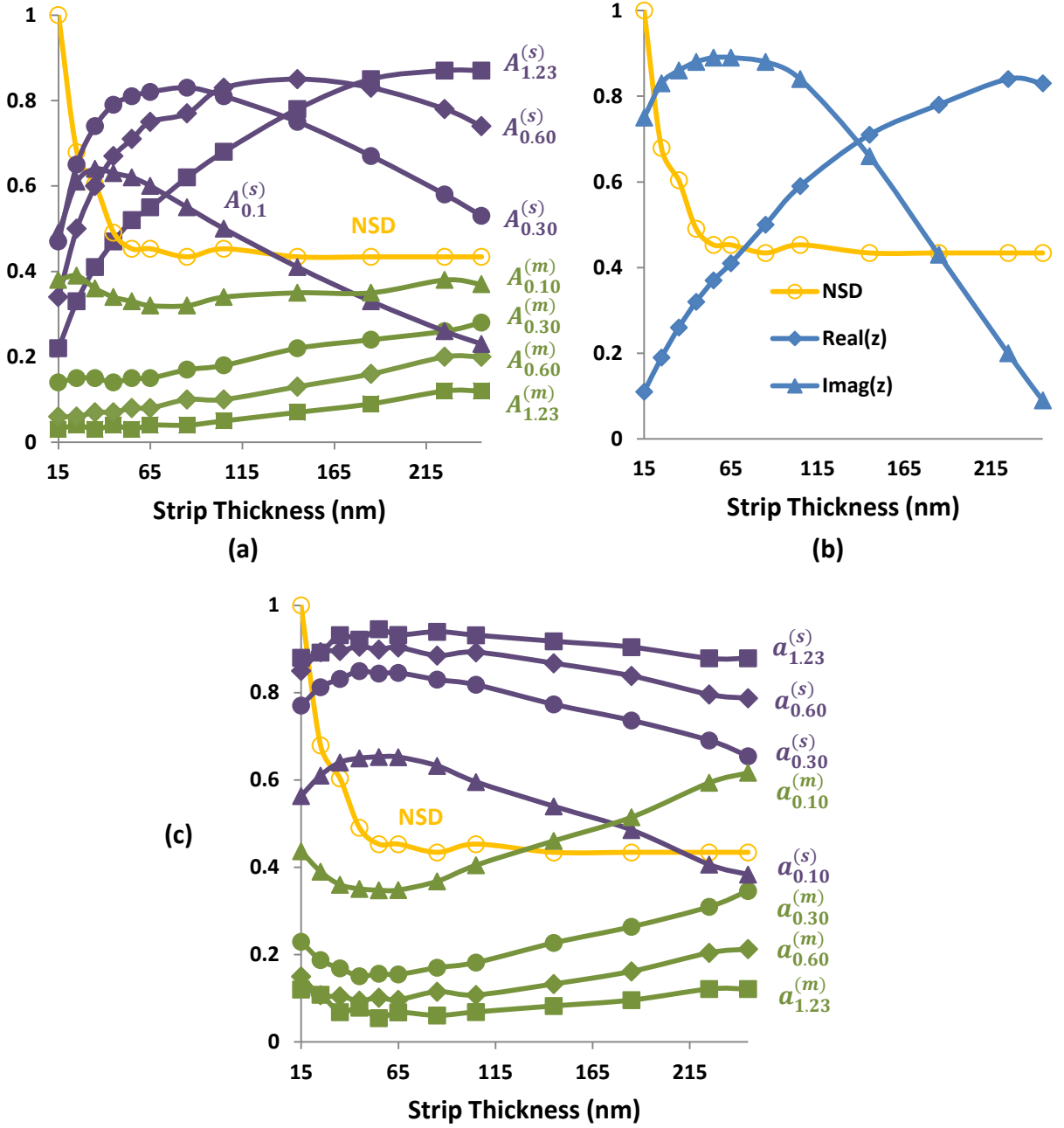

**Figure 4** (**a**) Absorbance and normalized skin depth (NSD) versus $t_s$. The purple curves show the peak absorbance in the spacer region. The green curves show the total absorbance in the metallic regions. Orange curve shows the skin depth normalized with respect to its maximum value of 53nm at $t_s = 15$nm. $A_j^{(i)}$ stands for absorbance in spacer ($i = s$) or metallic regions ($i = m$) for different $\Im(\varepsilon)$ values $j$ of the spacer. NSD is calculated at $\Im(\varepsilon) = 0$. $\Re(\varepsilon) = \Re(\varepsilon_1) = 0.86$ and $t_g = 245$nm. (**b**) Effective surface impedance of the metamaterial absorber and NSD [repeated from (a)] versus $t_s$ calculated near the total absorbance peak for $\varepsilon_1 = 0.86 + 1.23j$ and $t_g = 245$nm. (**c**) The percentage of total absorbance in different regions of the metamaterial absorber corresponding to (a) and NSD versus $t_s$. $a_j^{(i)}$ stands for the percentage of total absorbance in region $i$ for different $\Im(\varepsilon)$ values $j$ of the spacer.

In Fig. 4(a), for $t_s = 15$nm, we can rank the absorbance in the lossy spacer for different imaginary parts as $A^{(s)}_{0.1} > A^{(s)}_{0.3} > A^{(s)}_{0.6} > A^{(s)}_{1.23}$ where the superscript and subscript stands for the spacer and $\Im(\varepsilon)$, respectively. On the other hand, at $t_s = 35$nm, we observe that the absorbance starts to drop after a certain value of the $\Im(\varepsilon)$, that is $A^{(s)}_{0.3} > A^{(s)}_{0.1}$ but $A^{(s)}_{1.23} < A^{(s)}_{0.6} < A^{(s)}_{0.3}$. This means that the impedance mismatch produced by the reduced skin depth at $t_s = 35$nm is only to the extent that $\Im(\varepsilon) = 0.3$ and other larger imaginary values overcompensate the impedance mismatch and thus result in smaller absorbance. Therefore, the optimum value of $\Im(\varepsilon)$ should be somewhere between 0.3 and 0.6 to maximize the absorbance in the lossy spacer. This optimum value of the $\Im(\varepsilon)$ is the one which fully compensates the impedance mismatch opened by reduced skin depth at $t_s = 35$nm. Similarly, if we consider $t_s = 150$nm, the resultant total impedance mismatch should be compensated by an $\Im(\varepsilon)$ between 0.6 and $\Im(\varepsilon_1) = 1.23$. Since the impedance mismatch increases with $t_s$, the optimum value of $\Im(\varepsilon)$ also increases. We noticed that our observations here are consistent with a recent study by Hu, et al.[33] where they study the role of $\Im(\varepsilon)$ of the spacer on the impedance match and absorption. What we add here is the effect of changing skin depth which is not considered in Ref. [33].

Another important observation in Fig. 4(a) is that the curves for $\Im(\varepsilon)$ corresponding to the absorbance in the lossy spacer first increase and then decrease except for $\Im(\varepsilon) = \Im(\varepsilon_1) = 1.23$. The increase is rapid and occurs in the regime where the skin depth and impedance change rapidly whereas the decrease is more gradual [no decrease at all for $\Im(\varepsilon) = \Im(\varepsilon_1) = 1.23$] and occurs in the regime where the skin depth and impedance tend to converge. The peaks for different imaginary parts show that the specific value of $\Im(\varepsilon)$ cannot compensate the impedance mismatch if $t_s$ is increased further beyond the peak, because this results in smaller skin depth and/or higher impedance mismatch which requires larger $\Im(\varepsilon)$ for full compensation of the resultant impedance mismatch. The peak for $\Im(\varepsilon) = \Im(\varepsilon_1) = 1.23$ occurs at $t_s = 245$nm and corresponds to the peak in Fig. 3(d).

Regarding metallic losses in Fig. 4(a) we observe in general that the metallic losses decrease with increasing $\Im(\varepsilon)$. However, the change in the skin depth and impedance has also an impact on the metallic losses. For example, consider $\Im(\varepsilon) = 0.1$. At $t_s = 15$nm the structure with the lossless spacer was an impedance matched perfect absorber and had the maximum skin depth. Under this condition, Fig. 4(a) shows that the incorporation of any lossy spacer with the given relatively large $\Im(\varepsilon)$ values increases the impedance mismatch. The smaller the $\Im(\varepsilon)$, the smaller is the impedance mismatch. Among the given $\Im(\varepsilon)$ values, $\Im(\varepsilon) = 0.1$ is the smallest. Therefore, the reduction in metallic losses is the smallest for $\Im(\varepsilon) = 0.1$, in turn followed by $\Im(\varepsilon) = 0.3$, $\Im(\varepsilon) = 0.6$, and $\Im(\varepsilon) = \Im(\varepsilon_1) = 1.23$. For $\Im(\varepsilon) = 0.1$, if we increase the $t_s$, the metallic losses gradually first decrease and then increase. The initial decrease happens, because in this regime the metallic losses rapidly decrease as a result of rapid decrease in the skin depth [see Figs. 2(a) and (b)]. Additionally, increase in the impedance mismatch with increasing $t_s$ in this regime also plays a role in reducing the metallic losses. However, at larger $t_s$ values the contribution of the $\Im(\varepsilon) = 0.1$ to compensate the resultant mismatch decreases. Therefore, the metallic losses at these $t_s$ values cannot be reduced further below their original values when the structure was having lossless spacer. This also applies to larger $\Im(\varepsilon)$ values, which also exhibit increasing metallic losses with increasing $t_s$. On the other hand, in contrast with $\Im(\varepsilon) = 0.1$, the larger $\Im(\varepsilon)$ values show relatively flat and low metallic absorbance at small $t_s$ values. Because, initially at $t_s = 15$nm, the structure has substantially large impedance mismatch for all the $\Im(\varepsilon)$ values. Therefore, the absorbance in both lossy spacer and metals starts from the lowest values at $t_s = 15$nm. Increasing the $t_s$ reduces the metallic losses due to reducing skin depth. However, simultaneously improved impedance match increases the metallic losses. This is why the skin depth balances the impedance match in the flat region for metallic losses in Fig. 4(a).

Finally, Fig. 4(c) shows how the percentage of total absorbance changes in the lossy spacer and metals with $t_s$. At $t_s = 15$nm, if the structure would have lossless spacer, the metals would have 100% of the total absorbance whereas the spacer would have 0% [i.e., perfect absorber, Fig. 2(a)]. When we replace the lossless spacer with the lossy spacer with $\Im(\varepsilon) = 0.1$, we observe that the percentage of total absorbance in the lossy spacer (i.e., 56%) is higher than the metal (i.e., 44%). Higher percentages are obtained in the lossy spacer for larger $\Im(\varepsilon)$ despite smaller absolute absorbance due to decreasing impedance match at $t_s = 15$nm. As a result of overall interaction between lossy spacer and metals summarized in Fig. 4(a), we observe an interesting exchange of absorbance localization between the lossy spacer and metals as we increase the $t_s$ in Fig. 4(c). The percentage of total absorbance in the lossy spacer (metals) first increases (decreases) in the region where the skin depth rapidly changes and then decreases (increases) in the region where the skin depth converges.

**Scaling the Ohmic loss exchange toward the red edge of the visible spectrum.** To demonstrate the scalability of the photovoltaic absorber to Point 2 in Table 1, we start from a perfect absorber which is now tuned to a frequency around Point 2 by modifying the geometric parameters of the perfect absorber in Fig. 2(a). Considering the underlying resonant magnetic dipole mode displayed in Figs. 2(b) and (c), we can crudely estimate the resonance frequency from $\omega_0 \approx 1/\sqrt{LC}$ where L and C are equivalent circuit inductance and capacitance, respectively (More accurate expression requires detailed consideration of the mode distribution, but this does not change our conclusions drawn here). Because the $\Re(\varepsilon_2) = 4.59$ is substantially large compared to $\Re(\varepsilon_1) = 0.86$, scaling the resonant magnetic dipole mode to the neighborhood of Point 2 requires a sufficiently narrow strip width of 55nm to compensate increasing capacitance. The other geometric parameters mainly serve for optimization of the perfect absorber. The absorbance spectrum of the resultant perfect absorber is shown in Fig. 5(a). The trend we observe with the increasing metal thicknesses is similar to around Point 1 [see Figs. 5(a)-(d)]. This is due to the same electromagnetic mode tailored using BSDT. When the strip thickness gradually changes from 10nm to 135nm, the peak absorbance for the strip drops from above 75% to below 40%, while the absorbance in the ground plate, first, like before, increases to about 30% to accommodate photons not absorbed by the strip and then drops back to a level slightly above 20% with increasing thickness. Thus BSDT gives about 40% overall decrease in the metallic absorbance with lossless spacer. The magnetic field and current density distribution for the final absorber at the peak absorbance is shown in Fig. S4 (see Supplementary information). Despite thick metals the fields are tightly confined to the metal-semiconductor interfaces.

Restoring gradually the true value of $\Im(\varepsilon_2) = 0.45$ shifts most of the remaining metallic loss to the absorption in the semiconductor layer [see Figs. 5(e)-(g)]. The overall peak absorbance in the semiconductor layer reaches above 70% while the total metallic absorbance drop to below 30%. The maximum total absorbance in plain $In_{0.54}Ga_{0.46}N$ layer with the same thickness (i.e., $d = 12$nm) is only about 5% for the given frequency spectrum. Power loss density distribution in Fig. S5 (see Supplementary information) shows significant concentration of optical power in the semiconductor layer.

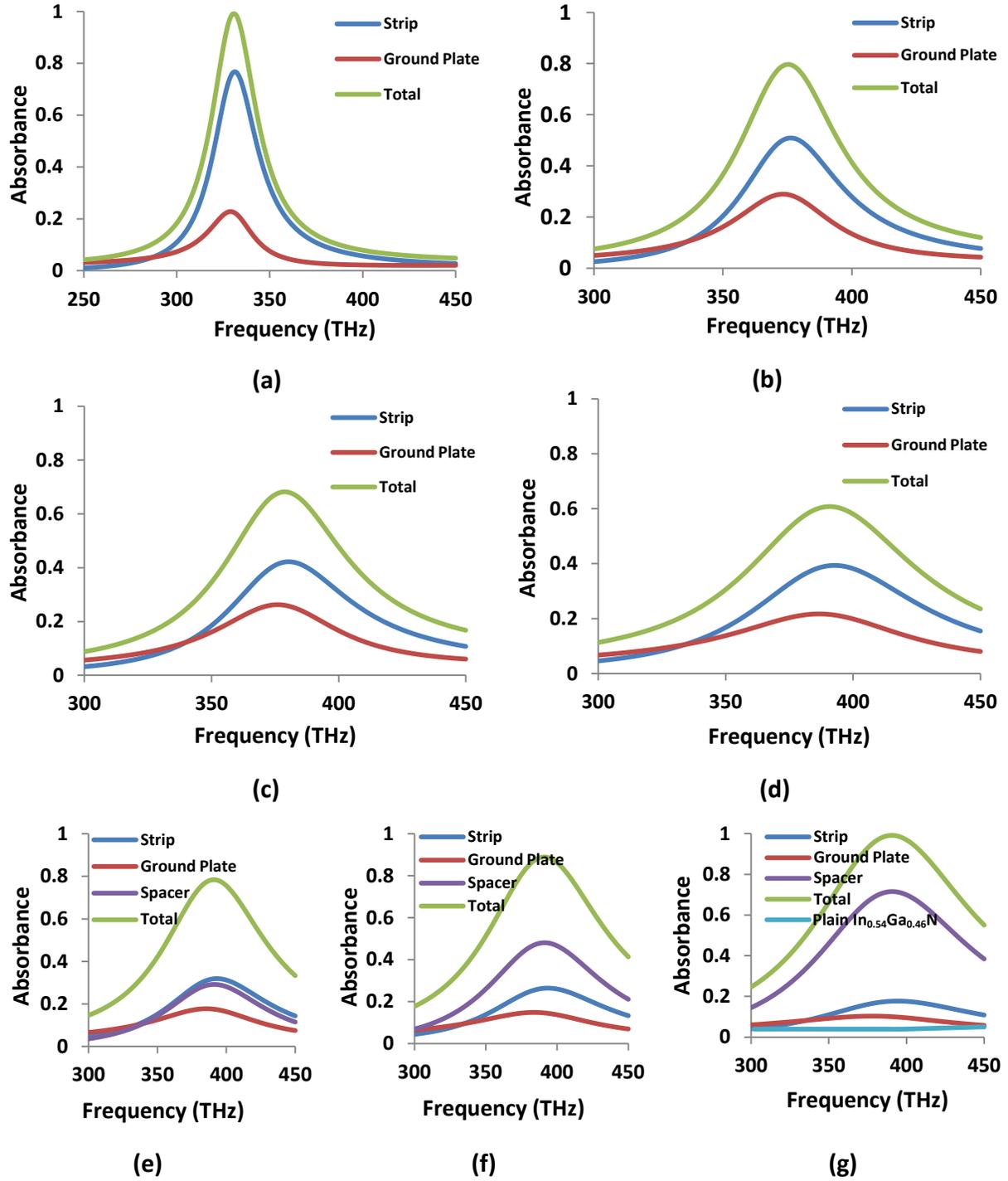

**Figure 5** Absorbance in different parts of the perfect absorber (**a**) operating in the neighborhood of Point 2, (**b**) when the strip thickness in the absorber increases to $t_s = 60$nm, (**c**) when both the strip and the ground plate have $t_s = t_g = 60$nm thickness, and (**d**) $t_s = t_g = 135$nm thickness. All the other parameters are kept fixed and the same as in Table 1. Absorbance in different parts of the absorber with the parameters same as in (d) except for (**e**) $\Im(\varepsilon) = 0.1$, (**f**) $\Im(\varepsilon) = 0.2$, and (**g**) $\Im(\varepsilon) = \Im(\varepsilon_2) = 0.45$. The total absorbance in plain In$_{0.54}$Ga$_{0.46}$N layer of the same thickness (i.e., $d = 12$nm) is also shown for comparison.

In comparison with Point 1, photovoltaic absorber at Point 2 has about 15% less peak absorbance in the semiconductor. This is mainly due to relatively small $\Im(\varepsilon_2) = 0.45$ compared to $\Im(\varepsilon_1) = 1.23$. Peak absorbance in the semiconductor layer should vary from about 85% to 70% between Point 1 and Point 2, respectively, due to the smooth variation of the complex permittivity between Point 1 and Point 2.

Below, similar to Point 1, we study the role of BSDT on the transition from perfect metallic absorbance in Fig. 5(a) to near-perfect semiconductor absorbance in Fig. 5(g). In Fig. 6(a) we study the absorbance characteristics of the structure around Point 2. Fig. 6(a) is obtained by replacing the lossless spacer with the lossy spacer similar to Fig. 4(a) such that $\Re(\varepsilon) = \Re(\varepsilon_2) = 4.59$ and $\Im(\varepsilon)$ is assumed to have two different values which are $\Im(\varepsilon) = 0.1$ and $\Im(\varepsilon) = \Im(\varepsilon_2) = 0.45$. At $t_s = 10$nm, the structure with the lossless spacer is an impedance matched perfect absorber [see Fig. 5(a)]. Here, if the lossless spacer is replaced with a lossy spacer with sufficiently small $\Im(\varepsilon)$ (e.g., 0.1) the lossy spacer cannot open an impedance mismatch. The total absorbance is shared between the metals and lossy spacer. In this case, when $\Im(\varepsilon) = 0.1$ and $t_s = 10$nm, the metallic absorbance drops to 69% from 100% and the absorbance in the spacer increases to 28% from 0%. However, if the $\Im(\varepsilon)$ is sufficiently large [e.g., $\Im(\varepsilon_2) = 0.45$], the impedance mismatch occurs, metallic losses are further reduced to 27% and absorbance in the lossy spacer increases to 48%. Thus the total absorbance drops to 75% due to the impedance mismatch. If we would further increase the $\Im(\varepsilon)$, the situation would be similar to $t_s = 15$nm in Fig. 4(a) [i.e., the larger is the $\Im(\varepsilon)$, the higher is the impedance mismatch, the lower is the absorbance in both metals and lossy spacer.)] When we increase the $t_s$, the impedance mismatch for the structure with lossless spacer increases. Larger $\Im(\varepsilon)$ compensates the opened impedance mismatch better than small $\Im(\varepsilon)$ through absorbance in the lossy spacer. This is similar to our observation in Fig. 4(a). Fig. 6(a) also provides an interesting observation which does not occur in Fig. 4(a). At sufficiently small $\Im(\varepsilon)$ the impedance compensation is mainly achieved by absorbance in metals, because the $\Im(\varepsilon)$ is not sufficiently large to reduce the metallic losses. However, at larger $\Im(\varepsilon)$ values, the impedance compensation is achieved by absorbance in the lossy spacer. Another difference between Fig. 4(a) and Fig. 6(a) is that the absorbance curves tend to converge with the converging skin depth. Fig. S6 (see Supplementary information) shows that the trends for the percentage of total absorbance for both metals and lossy spacer in the regime where the skin depth rapidly changes are similar to Fig. 4(c), but the percentages converge with the converging skin depth consistent with the absorbance curves at this frequency range around Point 2.

Briefly, Figs. 4(a) and 6(a) show that there are different regimes depending on the value of the $\Im(\varepsilon)$ for a given magnetic dipole mode based absorber. If the $\Im(\varepsilon)$ is sufficiently small, the incorporation of the lossy spacer does not change the impedance, but the metallic losses are not reduced much either [see, for example, $t_s = 10$nm in Fig. 6(a)]. If the $\Im(\varepsilon)$ is gradually increased to the extent that the impedance mismatch is compensated, the metallic losses decrease and the absorbance in the lossy spacer increases [see, for example, $t_s = 145$nm in Fig. 4(a)]. If the $\Im(\varepsilon)$ is increased beyond the point that the impedance mismatch is overcompensated, both metallic losses and absorbance in the lossy spacer is reduced [see again $t_s = 145$nm in Fig. 4(a)]. The ideal condition to maximize the absorbance in the lossy spacer occurs in the intermediate regime. The BSDT provides a simple guiding mechanism to achieve this ideal condition by controlling the impedance prudently in a desired fashion.

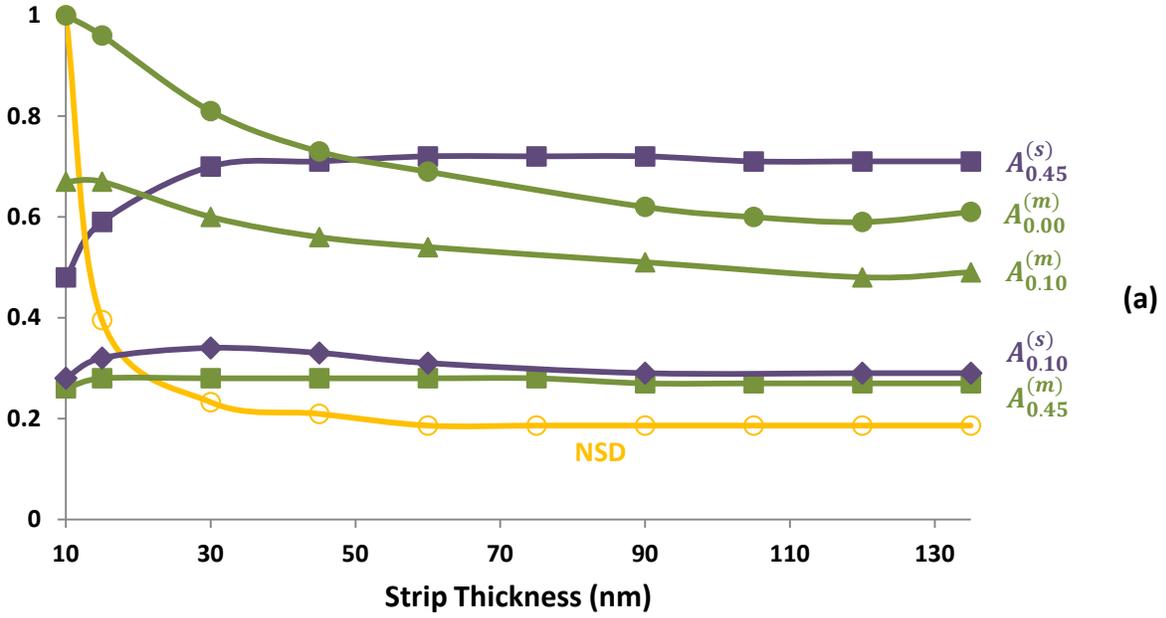

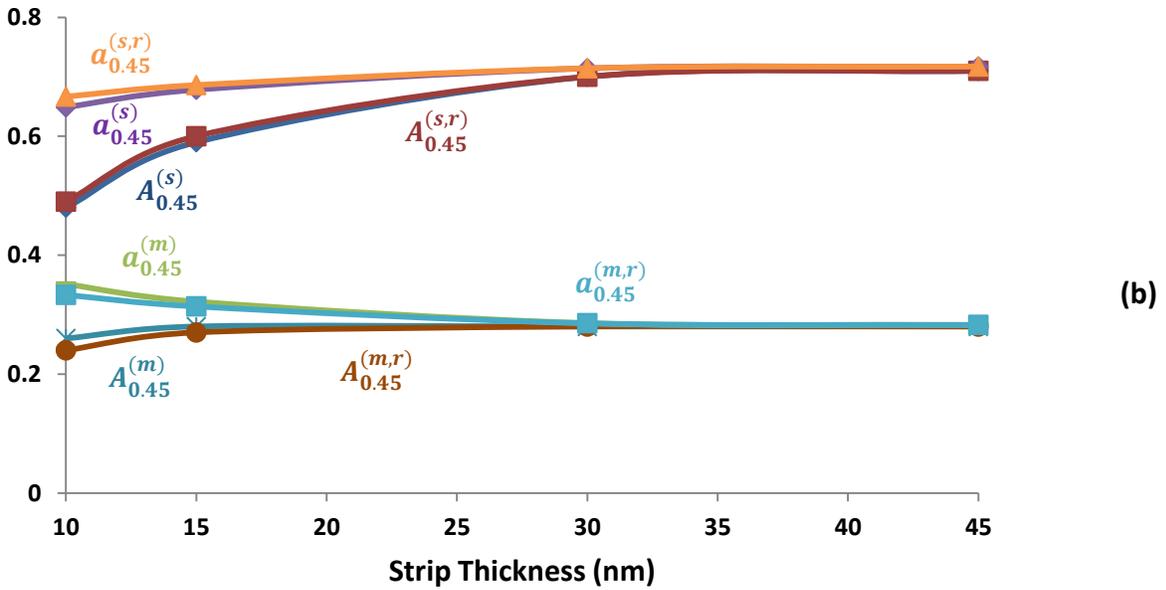

**Figure 6** (**a**) Absorbance and NSD versus $t_s$. The purple curves show the peak absorbance in the spacer region. The green curves show the total absorbance in the metallic regions. Orange curve shows the skin depth normalized with respect to its maximum value of 86nm at $t_s = 10$nm. $A_j^{(i)}$ stands for absorbance in region $i$ for different $\Im(\varepsilon)$ values $j$ of the spacer. NSD is calculated at $\Im(\varepsilon) = 0$. $\Re(\varepsilon) = \Re(\varepsilon_2) = 4.59$ and $t_g = 135$nm. (**b**) The absorbance $[A_j^{(i)}]$ and the percentage of total absorbance $[a_j^{(i)}]$ in different parts of the metamaterial absorber in (a) for $\Im(\varepsilon) = \Im(\varepsilon_2) = 0.45$ versus $t_s$ compared with the absorbance $[A_j^{(i,r)}]$ and the percentage of total absorbance $[a_j^{(i,r)}]$ of the metamaterial absorber with the constant resonance frequency at 378THz. The curves for $\Im(\varepsilon) = 0.1$ were not shown for clarity.

Finally, Fig. 6(b) shows that the shift in the resonance peaks due to the change in geometric or material parameters does not have important effect on the observed distributions of the power dissipation. To demonstrate this we kept the resonance frequency at 378THz [i.e., the resonance frequency when $t_s = 45$nm in Fig. 6(a)] for all the $t_s$ values by changing the original width (i.e., $w = 55$nm) of the top metal strip. The maximum shifts in the resonance frequency and $w$ from their original values occurred at $t_s = 15$nm. The shifts were 48THz and 10.3nm for the resonance frequency and $w$, respectively.

**Scaling the Ohmic loss exchange to the violet edge of the visible spectrum using higher order plasmonic modes.** Next, we scale the Ohmic loss exchange to Point 3, 695THz (i.e., violet wavelength). Despite relatively large $\Re(\varepsilon_3) = 5.67$, at this frequency, it is difficult to use the same magnetic dipole mode which requires vanishing strip width. Rather we can employ higher order plasmonic modes. Magnetic field distribution for one such mode is displayed in Fig. 7(a). This mode can be used to design a near-perfect absorber [i.e., above 95% peak absorbance, see Figs. 7(b)-(d)] in the neighborhood of Point 3 with a lossless spacer by tailoring the geometric parameters according to Table 1. Importantly, surface plasmons reside at the metal-semiconductor interfaces. This allows us to apply BSDT as before by simply increasing the metal thicknesses. Figs. 7(b)-(d) show total metallic peak absorbance drops down to about 40% by increasing the strip thickness from 20nm to 60nm (i.e., reaching the same thickness as the ground plate). Here, in contrast with Points 1 and 2, first, the peak absorbance in the ground plate of fixed thickness decreases with the increasing strip thickness due to the increasing impedance mismatch at the corresponding frequencies. Second, the metals do not have to be very thick. When we incorporate the true value $\Im(\varepsilon_3) = 1.61$ following the BSDT, the absorbance in the semiconductor layer at 695THz exceeds 85% [see Figs. 7(e)-(g)]. In contrast, the total absorbance in plain $In_{0.54}Ga_{0.46}N$ layer for the same thickness (i.e., $d = 31$nm) is between the ranges of $15\% - 27\%$. Thus, we convert Ohmic loss dominated near-perfect absorber to near-perfect photovoltaic absorber. The power loss density distribution in Fig. 7(h) shows high concentration of absorbed power in the semiconductor layer in consistency with Fig. 7(g).

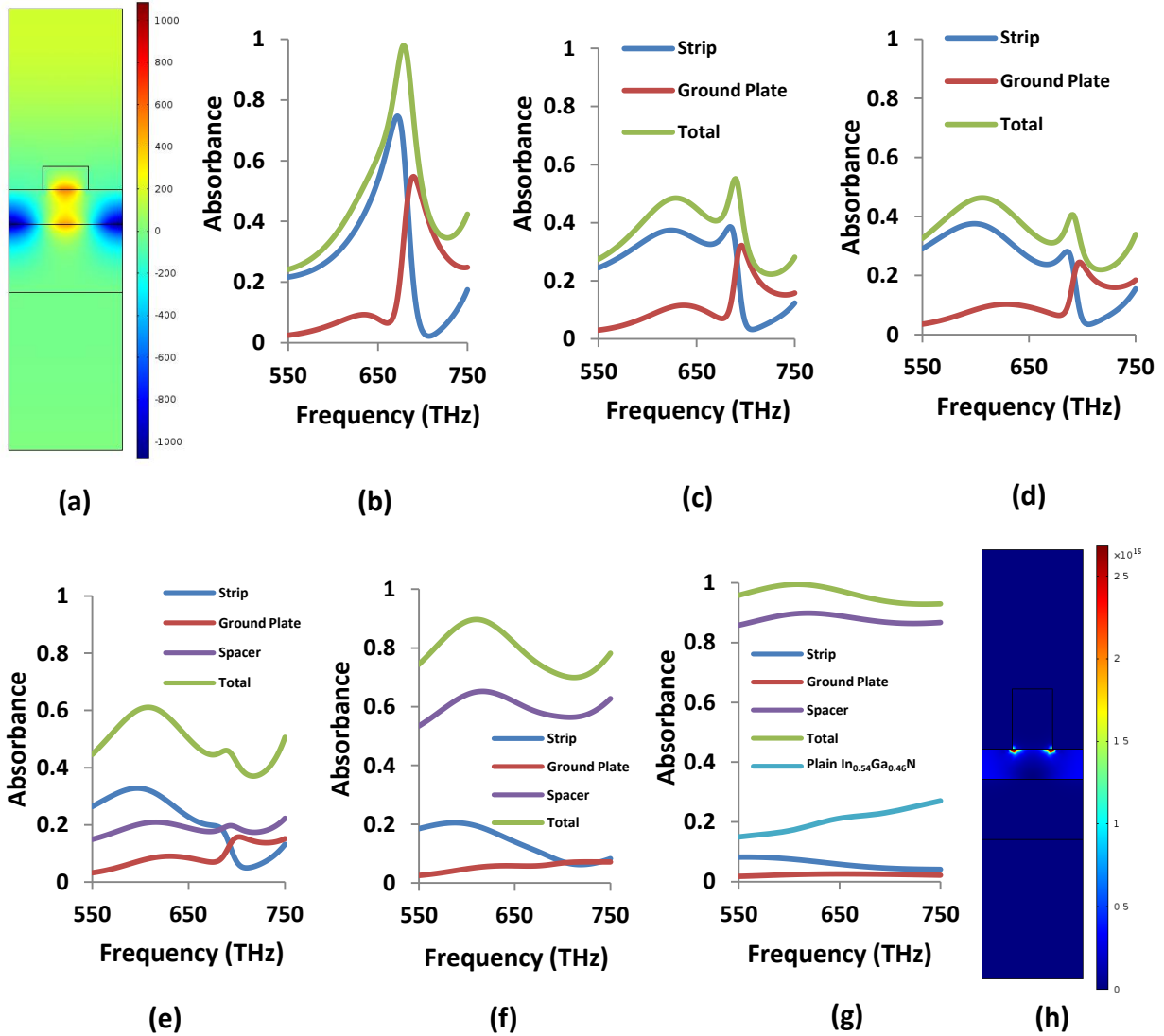

**Figure 7** (**a**) $H_z$ (A/m) for a higher order plasmonic mode resonating at 672THz (i.e., in the neighborhood of Point 3) supported by the near-perfect absorber with the parameters given in Table 1. Absorbance in different parts of the perfect absorber (**b**) operating in the neighborhood of Point 3 (legends not shown for clarity, same as other plots), (**c**) when the strip thickness in the absorber increases to $t_s = 40$nm, and (**d**) when both the strip and the ground plate have $t_s = t_g = 60$nm thickness. All the other parameters are kept fixed and the same as in Table 1. Absorbance in different parts of the absorber achieved by the BSDT process described in (b)-(d) followed by the gradual incorporation of the true complex permittivity such that (**e**) $\Im(\varepsilon) = 0.1$, (**f**) $\Im(\varepsilon) = 0.5$, and (**g**) $\Im(\varepsilon) = \Im(\varepsilon_3) = 1.61$. The total absorbance in plain $In_{0.54}Ga_{0.46}N$ layer of the same thickness (i.e., $d = 31$nm) is also shown for comparison. (**h**) Power loss density (W/m³) distribution for the near-perfect absorber in (g) for the third selected frequency of 695THz in the $In_{0.54}Ga_{0.46}N$ data (i.e., Point 3 in Table 1). Significant portion of the power (i.e., over 85%) is absorbed inside the $In_{0.54}Ga_{0.46}N$ layer.

**Ohmic loss exchange at the ultraviolet frequencies.** Scaling the absorber to the UV frequencies at Points 4 and 5 requires employing yet higher order surface plasmon modes with the given structure. However, this makes the BSDT less effective due to the complexity of the modes as explained below. Nevertheless, high absorbance of about 95% can be still achieved in the semiconductor layers mainly due to the large imaginary part of the permittivity of $In_{0.54}Ga_{0.46}N$. The detailed discussion for the UV frequencies is given in the Supplementary information.

**Discussion**

Based on simple metamaterial perfect (or near-perfect) absorber designs scalable to any frequency in the solar spectrum, seemingly contrary to common sense, we have shown that metallic layers in the absorbers do not necessarily constitute a problem for photovoltaic applications. The BSDT along with the bulk absorbance characteristics of the semiconductors, associated with the imaginary part of their complex permittivity, can allow for exchanging undesired resistive losses with the useful optical absorbance in the semiconductor layers. Thus, Ohmic loss dominated perfect (or near-perfect) absorbers can be converted into photovoltaic near-perfect absorbers with the advantage of harvesting full potential of light management capability offered by perfect absorber designs. It is also worth mentioning that our results confirm the recent work in Ref. **34**.

The BSDT provides a simple guiding mechanism to control the impedance prudently in a desired fashion. If there is no such a guiding mechanism such as BSDT, with a capability alone to simply tune the impedance, it is difficult to control and predict in an effective manner where the absorbance is localized in the structure. In fact, this is the same reason why the BSDT as applied to some of the higher order modes (see, for example, Fig. S9, Supplementary information) is less effective; because the impedance is only partially controlled by the BSDT due to more complicated mode profiles of the higher order modes which are not fully accounted by the BSDT. For simplicity, the way we have applied the BSDT considers only the surface waves on the metal-spacer interface. However, the BSDT can be applied more effectively by considering all the existing surface waves constituting the mode.

The perfect absorption in earlier reports[35] on metamaterial absorbers was qualitatively attributed to matching the impedance $z = \sqrt{(\mu_{eff}/\varepsilon_{eff})}$ of the bulk metamaterial to that of vacuum where $\mu_{eff}$ and $\varepsilon_{eff}$ are the bulk effective magnetic permeability and bulk effective electric permittivity of the metamaterial, respectively. However, this viewpoint has not received wide acceptance for the conceptual reasons as discussed below. Instead, the metamaterial perfect absorbers have been described more meaningfully by several different models such as single-resonance[15, 32], perfectly impedance-matched sheet[36], and equivalent circuit[32, 36, 37] models. In these models, the metamaterial absorbers are defined directly by their effective impedance to avoid the ambiguity in defining $\mu_{eff}$ and $\varepsilon_{eff}$, since the metamaterial perfect absorber cannot be strictly considered homogeneous bulk media. To resolve the ambiguity, in a different model[38] the whole metamaterial was considered as a three-layer weakly coupled system consisting of two metamaterials separated by a dielectric spacer and an interference theory[37-41] was used to explain the perfect absorption. Non-zero order Bragg waves supported by the metamaterial perfect absorbers were treated in a grating theory[42] where the periodicity of the metamaterial was rigorously taken into account.

The metamaterial perfect absorber structures with geometry similar to ours have been modeled under a single-resonance approximation in Refs. **15** and **39**. In this model, an effective surface impedance of the metamaterial structure is calculated. The perfect absorption occurs when the effective surface impedance of the metamaterial structure matches to that of vacuum. Single-resonance model can be readily applied to our metamaterial absorbers operating in the frequency spectrum between Points 1 and 2 since we use exactly the same magnetic dipole mode as in Refs. **15** and **39**. Therefore, the single-resonance model was used in the calculation of the effective surface impedance shown in Fig. 4(b) to explain the effect of BSDT on localization of absorption in different parts of the metamaterial absorber. Although it is not necessary for the scope of the paper, the metamaterial absorbers operating around Points 3-5 can be also described by an appropriate model considering the underlying resonant modes.

BSDT, as applied here, becomes especially effective in the large part of the solar spectrum below UV frequencies. Around UV frequencies, bulk absorbance characteristics of the semiconductor dominate the change in absorption localization possible with BSDT alone. Based on experimentally measured $In_{0.54}Ga_{0.46}N$ data and a "conservative" Drude model for gold[43], we have shown that between $75\% - 95\%$ optical absorbance can be achieved in the semiconductor layers of the converted absorbers depending on the targeted frequency in the solar spectrum. However, considering more realistic permittivity data for gold including the size effects may result in slightly less shift of the absorption localization to the semiconductor layers than predicted here especially at small solar wavelengths.

It is desired that optical enhancement in absorbance in the active region of the solar cell is maximized. This can be best achieved by increasing absorbance in the entire solar spectrum. Thus, the larger the operating bandwidth of the metamaterial photovoltaic absorber, the more efficient will be the solar cell. However, even a narrowband absorber operating around a specific frequency in the solar spectrum can provide overall optical enhancement in the solar spectrum compared to reference solar cell (see, for example, Ref. **27**); because the optical enhancement in absorbance achieved near the operating frequency (or frequencies if multiple resonances of a single absorber is considered) of the absorber is usually sufficient to obtain important overall enhancement in the entire solar spectrum. This is true even if the metamaterial structure does not perform as a perfect absorber. Here, we purposely tune the structure as a near-perfect photovoltaic absorber using BSDT. Therefore, overall optical enhancement should be more optimal for the near-perfect photovoltaic metamaterial absorbers guided by the BSDT.

We should also note that solar cell architectures where the metals are completely buried inside the semiconductor can provide more efficient conversion process, because this increases the interaction surface between the metal and the semiconductor and enable more surface fields to interact with the electrons in the semiconductor. This is likely for $In_xGa_{1-x}N$ PV consisting of multiple junctions, which would be necessary to obtain the highest potential efficiencies for this material system. This becomes especially important when more complex higher order surface plasmons are exploited. Although we chose gold and $In_xGa_{1-x}N$ material system as an example to illustrate the concept of absorption localization change here, we can use any material systems including metals (silver, copper, aluminum, etc.), heavily doped semiconductors, nitrides[44], oxides[44,45], amorphous silicon, graphene and other organic materials provided that they have compatible fabrication process and engineered to support surface plasmons. Our approach can be also extended to broadband absorbers[14-19], super absorbers[19], photodetectors, metamaterials[24,25,46], plasmonic and other metal and semiconductor based optical devices where resistive losses and power consumption are important pertaining to the device performance. In addition, this approach can be applied to optimize the optical enhancement of solar PV devices of any arbitrary incoming spectrum. For example, recent work has shown that the spectral albedo could play a significant role in the optimization of PV systems designed in snowy regions[47]. Following this work metamaterials can be designed to tailor PV devices for specific environmental conditions resulting in non-standard solar spectra (e.g. heavy pollution of a high UV index).

**Methods**

**Simulation.** All the structures shown here are designed and simulated using fully-vectorial finite element based commercial software package COMSOL. Gold is modeled by conservative Drude model using plasma frequency $f_p = 2100$THz, and collision frequency $f_c = 19$THz[15]. Due to the periodic arrangement of the plasmonic nanostructures, periodic boundary conditions are used in the simulations for the vertical boundaries and ports are used for the top and bottom boundaries. The ports were placed in air and sufficiently away from the metamaterial absorbers to avoid the evanescent fields. The distance between the input port (output port) and the top (bottom) surface of the metal strip (ground plate) was chosen 140nm for the Points 1-3 and 200nm for the Points 4 and 5. Normally incident plane wave illumination with horizontal polarization (i.e., parallel to the metal-spacer interface) and 1W input power, constant over the frequency spectrum, was used for the excitation. The power absorbed in specific regions of the metamaterial absorber was calculated by integrating the time-averaged power loss density [i.e., $\frac{1}{2}\Re(\mathbf{E}.\mathbf{J}^*)$ where **E** and **J** are complex electric field intensity and current density, respectively] over the specified regions. We also verified the results independently from power flow and s-parameters. The absorbance was calculated by normalizing the absorbed power in specific parts of the absorber with respect to input constant power. We did not use any reference solar spectral irradiance, since no specific solar cell architecture is considered here. Near Points 4 and 5, the diffraction has contribution to the calculated absorbance spectra. However, this does not invalidate the detailed qualitative discussion in the Supplementary information on absorbance localization but provides useful insight about the role of BSDT on complex high order modes and how these modes distribute the power inside the structure.

**Acknowledgement**

This work was supported by the National Science Foundation under grant CBET-1235750 and in part under grant ECCS-1202443 and by the Oak Ridge Associated Universities. We would like to thank Koray Aydin at Northwestern University for discussion on impact of the proposed technique for solar PV.

**Author Contributions**

D. Ö. G. conceived the idea of shifting the Ohmic loss from metals to semiconductors using BSDT. D. Ö. G. and A. V. performed numerical analysis. D. Ö. G., J. M. P., and A. K. supervised the project. A. V., J. G., N. P., A. K., J. M. P., and D. Ö. G. co-wrote the manuscript.

**Additional Information**

**Competing financial interests:** The authors declare no competing financial interests.

# Supplementary Information: Exchanging Ohmic Losses in Metamaterial Absorbers with Useful Optical Absorption for Photovoltaics

Ankit Vora[1], Jephias Gwamuri[2], Nezih Pala[3], Anand Kulkarni[1], Joshua M. Pearce[1,2], and Durdu Ö. Güney[1,*]

[1]Department of Electrical and Computer Engineering, Michigan Technological University, Houghton, MI 49931, USA
[2]Department of Materials Science and Engineering, Michigan Technological University, Houghton, MI 49931, USA
[3]Department of Electrical and Computer Engineering, Florida International University, Miami, FL 33174 USA
*Corresponding Author: dguney@mtu.edu

**Previous applications[1,2] of the BSDT.**

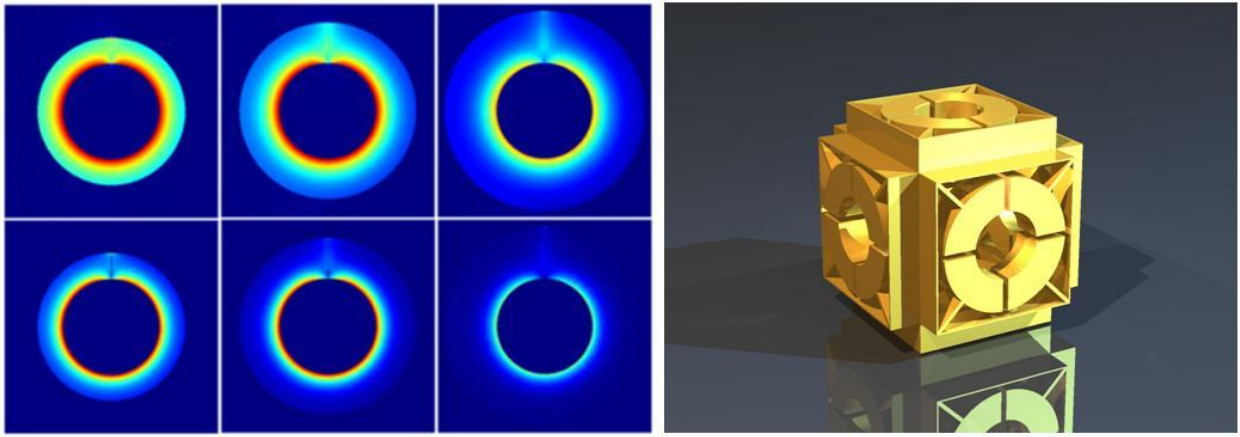

(a)  (b)

**Figure S1** (**a**) BSDT can be used to significantly reduce the geometric skin depth by increasing the radius of curvature of the surfaces where the resonant current flows. Thus the resonant current cannot penetrate deep into the metal; rather it is tightly bound to the surface. The top (bottom) panel from left to right shows the current (power loss) density distribution for split-ring-resonators with increasing radius of curvature which is achieved by making the rings thicker and bulkier. Only two-dimensional cross-sections are shown. The current density ranges between 0 (dark blue) and $6\times10^{14}$ A/m$^2$ (dark red) and the power loss density ranges between 0 (dark blue) and $4.5\times10^{21}$ W/m$^3$ (dark red). [modified from Figure 6 in "Guney, D. O., Koschny, T. & Soukoulis, C. M. Reducing ohmic losses in metamaterials by geometric tailoring. *Phys. Rev. B* **80**, 125129 (2009)"] (**b**) This effect can be further employed to design a fully connected three-dimensionally isotropic negative index metamaterial at optical frequencies by assembling the split-ring-resonators in the form of a cubic unit cell. The electrical interconnects do not short circuit the split-ring resonators due to the tight confinement of the loop currents. [Figure 2(b) in D. O. Guney, Th. Koschny, and C. M. Soukoulis, Intra-connected three-dimensionally isotropic bulk negative index photonic metamaterial, *Opt. Express* **18**, 12348 (2010).]

**Metamaterial absorbers with $In_xGa_{1-x}N$ spacers.**

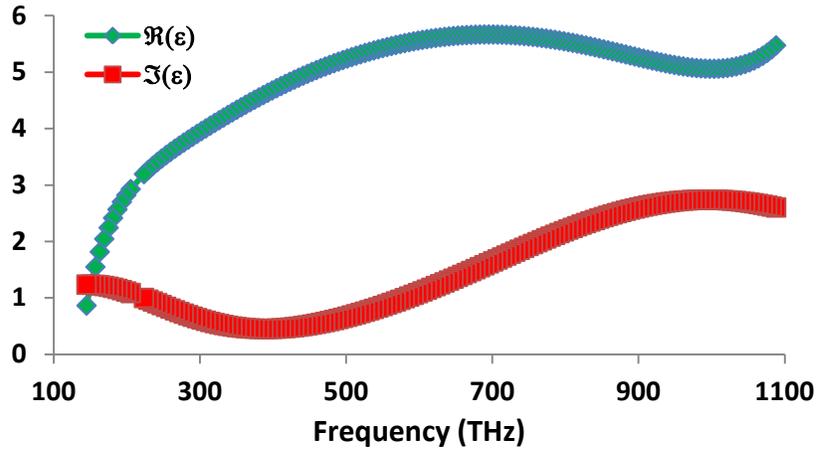

**Figure S2** Experimental complex permittivity data for $In_{0.54}Ga_{0.46}N$.

**Exchanging Ohmic losses at the infrared spectrum.**

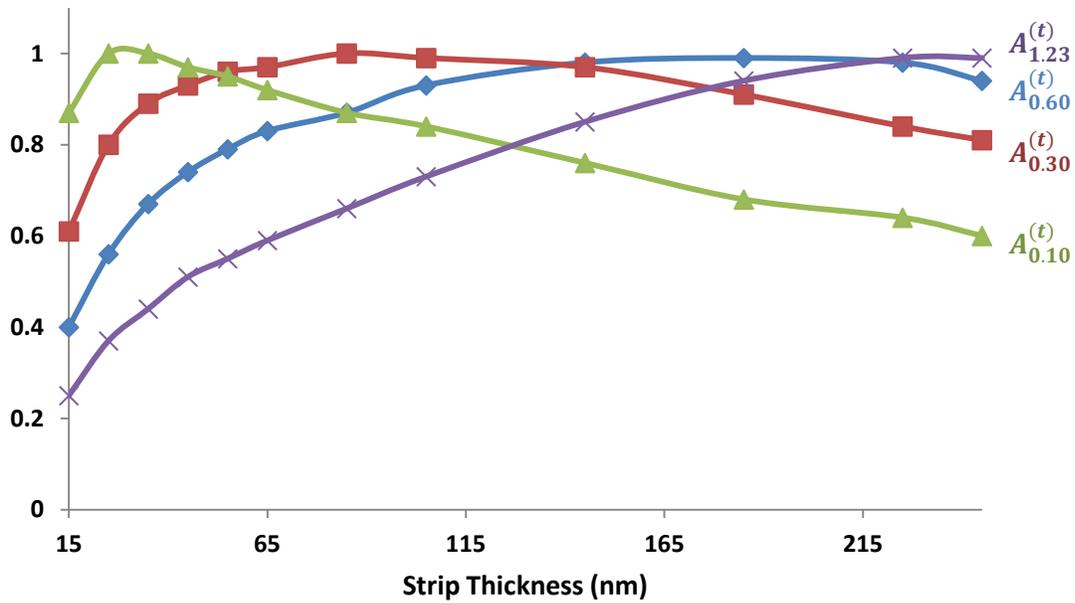

**Figure S3** Total absorbance (i.e., sum of the absorbance in the spacer and metallic regions) $A_j^{(t)}$ for different $\Im(\varepsilon)$ values $j$ of the spacer versus $t_s$ corresponding to Fig. 4(a).

**Scaling the Ohmic loss exchange toward the red edge of the visible spectrum.**

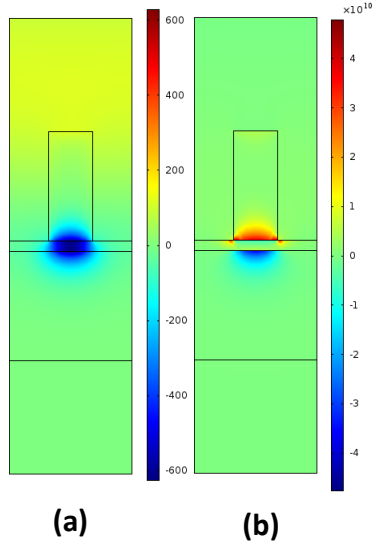

(a)     (b)

**Figure S4** (**a**) $H_z$ (A/m), and (**b**) $J_x$ (A/m$^2$) for the absorber in Fig. 5(d) at the peak total absorbance.

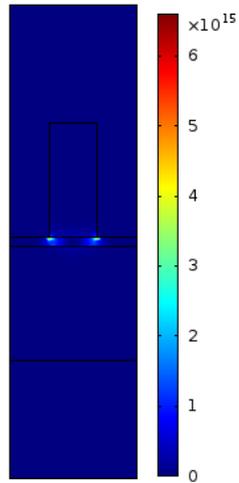

**Figure S5** Power loss density (W/m$^3$) distribution for the near-perfect absorber in Fig. 5(g) for the second selected frequency of 387THz in the In$_{0.54}$Ga$_{0.46}$N data (i.e., Point 2 in Table 1). Significant portion of the power (i.e., over 70%) is absorbed inside the In$_{0.54}$Ga$_{0.46}$N layer.

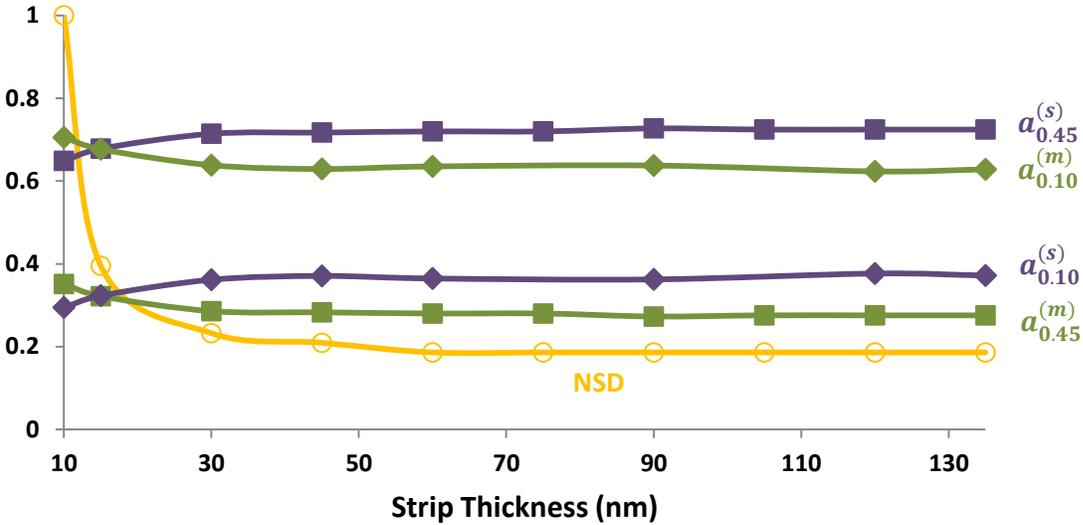

**Figure S6** The percentage of total absorbance $a_j^{(i)}$ in different regions of the metamaterial absorber corresponding to Fig. 6(a) and NSD versus $t_s$.

**Ohmic loss exchange at the ultraviolet frequencies.** The resonant peaks in the neighborhood of Point 4, 1000THz, are displayed in Fig. S7. We observe four resonant peaks, one of which has near-perfect absorption at 950THz. Magnetic field distributions for these resonance peaks from left to right are shown in Figs. S8(a)-(d), respectively. We will call these modes as Modes I-IV. All the resonant modes except the near-perfect absorber mode (i.e., Mode II) have significant reflectance. The Modes I and IV have additionally significant transmittance, which can be minimized by increasing the thickness of the ground plate. Thus, the part of the transmitted power can be converted to absorbed power.

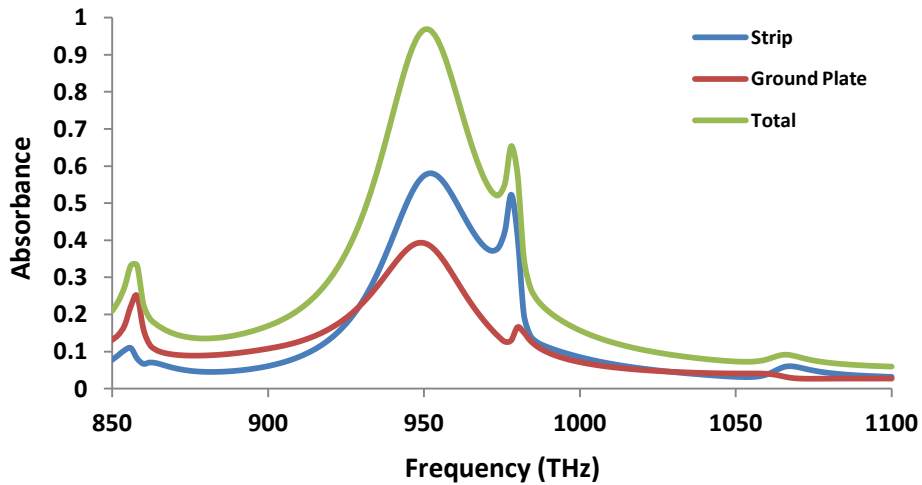

**Figure S7** Absorbance spectra in the neighborhood of Point 4 for different parts of the absorber. The second resonance at 950THz gives near-perfect absorbance. All the geometric parameters are listed in Table 1.

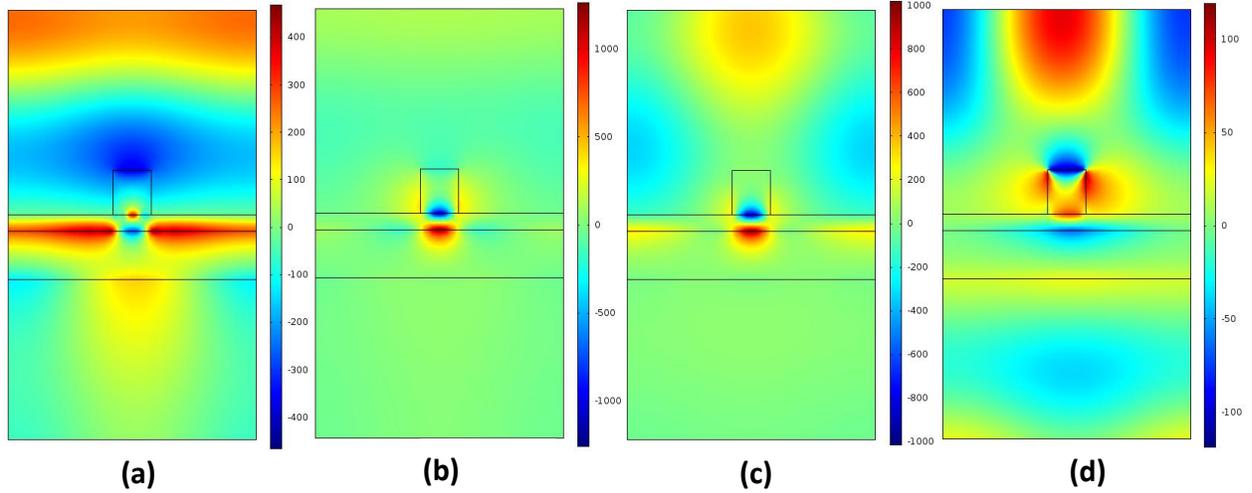

**Figure S8** $H_z$ (A/m) for (**a**) Mode I, (**b**) Mode II, (**c**) Mode III, and (**d**) Mode IV, corresponding to the four resonant peaks, respectively, in Fig. S7.

Unlike previous Points 1-3, we notice all the modes in Figs. S8(a)-(d) have field confinement at both metal-spacer and metal-air interfaces. However, to use BSDT more effectively for exchanging the metallic losses with optical absorption in the semiconductor layers, surface plasmons are desired to reside at only metal-semiconductor interfaces, so that they can be pushed further toward the semiconductor surface. Fortunately, here, large $\Im(\varepsilon_4)$ help achieve significant absorbance in the semiconductor layer despite surface plasmons residing at the metal-air interfaces. In general, interfacing metals with other lossy materials, especially if surface plasmons reside at those interfaces, should be avoided to maximize the absorption localization change from metals to semiconductor layers.

Figs. S7 and S9 show how BSDT, as applied to metal-spacer interface only, affects the metallic losses for each resonant mode. For Mode I [see Fig. S8(a)], increasing the metallic thicknesses gives a small decrease in the metallic loss from near 35% to below 25%. Even if we do not start from a perfect absorber (i.e., there exists transmittance and reflectance) and there are surface plasmons residing at metal-air interfaces, BSDT still gives reduction in metallic losses. Because the surface plasmons mainly reside at the interfaces [i.e., interfaces parallel to the metal-spacer interfaces, which asymptotically approach to the surface of a bulk (or semi-infinite) metal under the way we apply the BSDT], which comply with the BSDT. The near-perfect absorber mode, Mode II [see Fig. S8(b)], supports also weak surface plasmon oscillations at the vertical metal-air interfaces which are not accounted by BSDT. Therefore, this limits the reduction in the metal regions to about 70% absorbance (i.e., about 26% decrease in metallic losses). In comparison with the perfect (or near-perfect) absorbers in Points 1-3, this mode gives the least reduction in the metallic absorbance using BSDT.

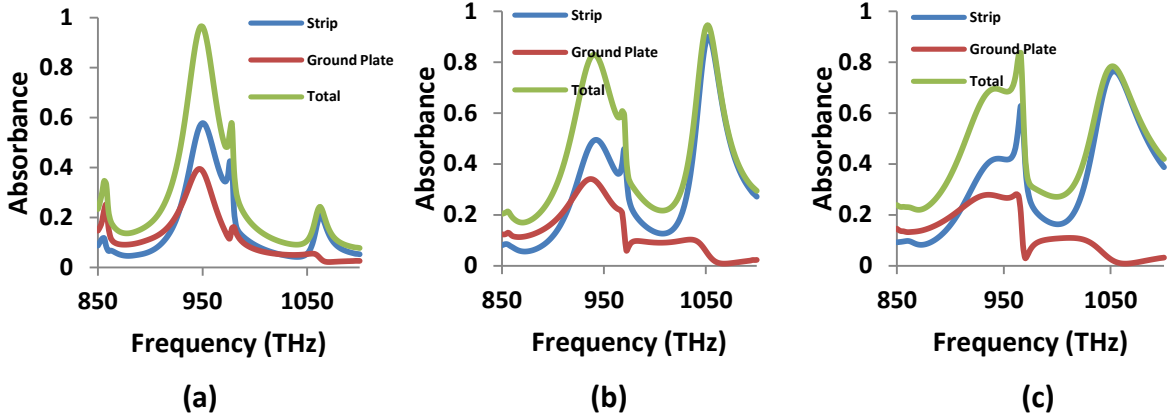

**Figure S9** Absorbance spectra in the neighborhood of Point 4 when (**a**) $t_s = t_g = 60$nm, (**b**) $t_s = t_g = 90$nm, and (**c**) $t_s = t_g = 105$nm. All the other parameters are kept fixed and the same as in Table 1.

The absorbance trend for Mode III [see Fig. S8(c)] is quite different than previous modes. The metallic absorbance first decreases from above 65% at 55nm strip thickness to about 55% at 60nm thickness. Then, the absorbance reaches above 80% when the metal thicknesses become 105nm. In contrast with all the other previous modes, here, the absorbance trend is not monotonic. Because when the metal thicknesses approach 105nm, a new resonance starts to emerge at about the same frequency as Mode III. This new mode can be distinguished from Mode III by carefully comparing Fig. S8(c) with Fig. S8(b) and Fig. S10. Notice that Fig. S8(c) has the reminiscence of Mode II in Fig. S8(b) due to the close proximity. Moreover, the magnetic field distribution at the horizontal strip-air interface in Fig. S10 manifests a strong surface wave which does not exist in Fig. S8(c). To distinguish this new mode from other modes, we will call Mode V. The non-monotonic trend that we see here shows that BSDT does not always result in reduced losses, especially when it leads to a resonance condition. Such a non-monotonic trend is also observed for Mode IV [see Figs. S7, S8(d) and S9]. The metallic losses increase until near-perfect absorbance is reached at 90nm metal thicknesses. Above this thickness, BSDT gives reduction in the metallic losses from about 95% to below 80%. Similar to Mode II, the existence of surface waves at the vertical metal-air interfaces limit the effectiveness of BSDT for Mode IV.

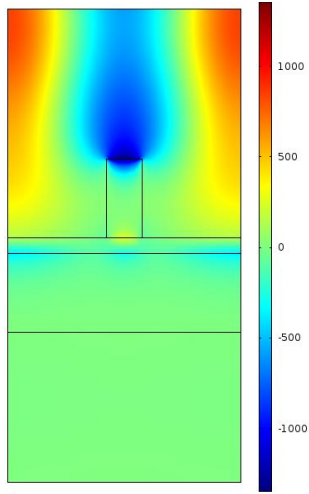

**Figure S10** $H_z$ (A/m) for Mode V at 966THz (i.e., absorbance peak for the strip) in Fig. S9(c).

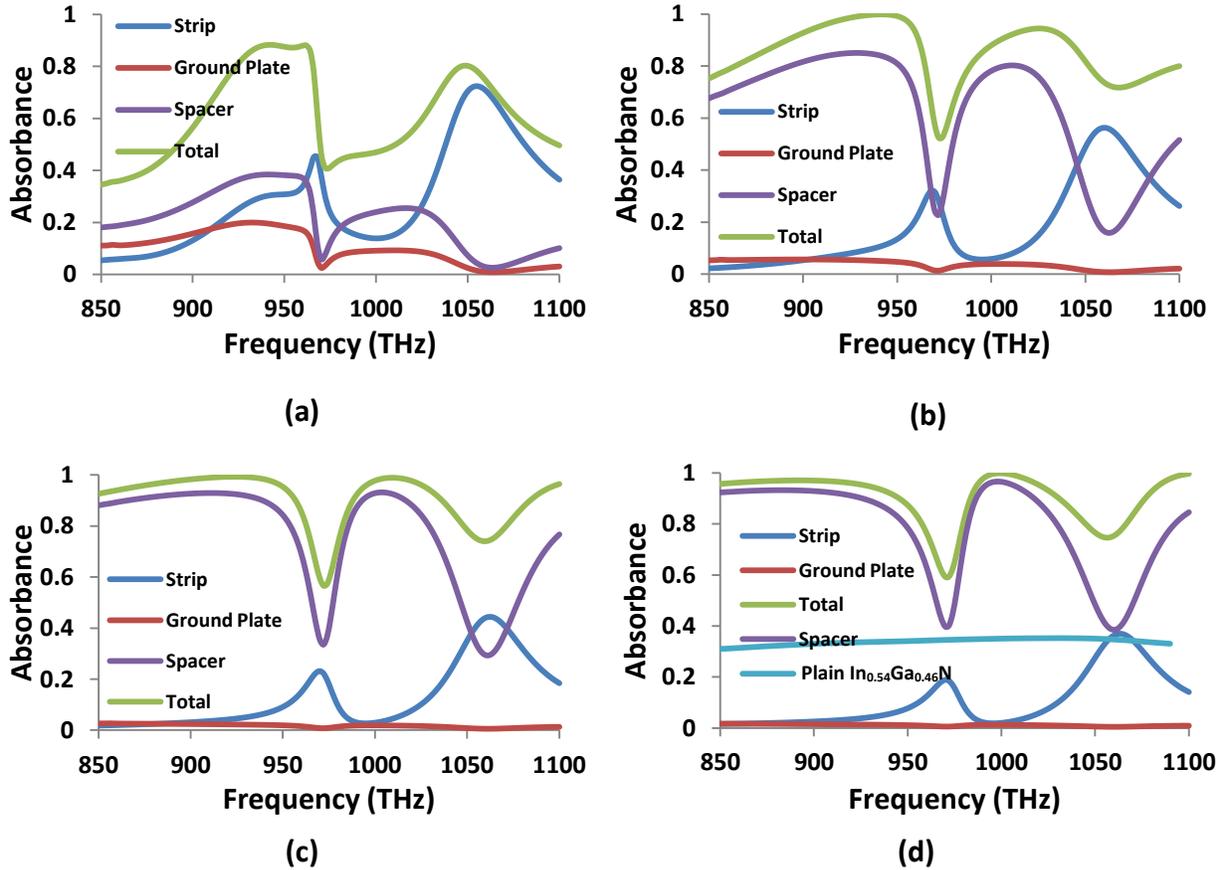

**Figure S11** Absorbance in different parts of the absorber achieved by the BSDT process described in Fig. S9 followed by the gradual incorporation of the true complex permittivity such that (**a**) $\Im(\varepsilon) = 0.1$, (**b**) $\Im(\varepsilon) = 0.74$, (**c**) $\Im(\varepsilon) = 1.74$, and (**d**) $\Im(\varepsilon) = \Im(\varepsilon_4) = 2.74$. The total absorbance in plain $In_{0.54}Ga_{0.46}N$ layer of the same thickness (i.e., $d = 20nm$) is also shown for comparison based on the experimental complex permittivity data in Fig. S2.

Having applied the BSDT, Fig. S11 shows how the restoring $\Im(\varepsilon_4) = 2.74$ affects the absorbance in the spacer. Fig. S12 shows the magnetic field distribution for the three absorbance peaks in Fig. S11(a) corresponding to the strips. Comparing these field distributions with Figs. S8(b), S10, and S8(d), respectively, and investigating Fig. S11, we observe that Modes IV and V are the only modes which survive the imaginary part of the permittivity of the semiconductor. Although Mode II persists at low imaginary part, it cannot survive the true value of the $\Im(\varepsilon_4)$. This is because, if we carefully investigate the field distribution for the Modes IV and V, they have significant portion of the field enhancement at metal-air interfaces compared to Mode II, where most of the field enhancement is at the metal-spacer interface. Similarly, Mode I should also have most of the field enhancement at the metal-semiconductor interface as can be seen from Fig. S8(a). Therefore, restoring the imaginary part of the semiconductor permittivity to its true value increases absorbance significantly in the neighborhood of Point 4 except in the frequency regions which mainly correspond to Modes IV and V. The absorbance in the semiconductor layer at Point 4 exceeds 95%. The absorbance in the semiconductor layer at the frequency position, where originally the near-perfect absorber mode (i.e., Mode II) existed, reaches about 80%. In contrast, the total absorbance in plain $In_{0.54}Ga_{0.46}N$ layer for the same thickness (i.e., $d = 20$nm) is between the ranges of $31\% - 35\%$ for the available experimental complex permittivity data in Fig. S2. Based on these important observations, we should also note that embedding metallic parts of the perfect absorbers in the semiconductors may help increase absorbance in the semiconductor by eliminating undesired surface waves at metal-air interfaces especially when higher order modes are of concern.

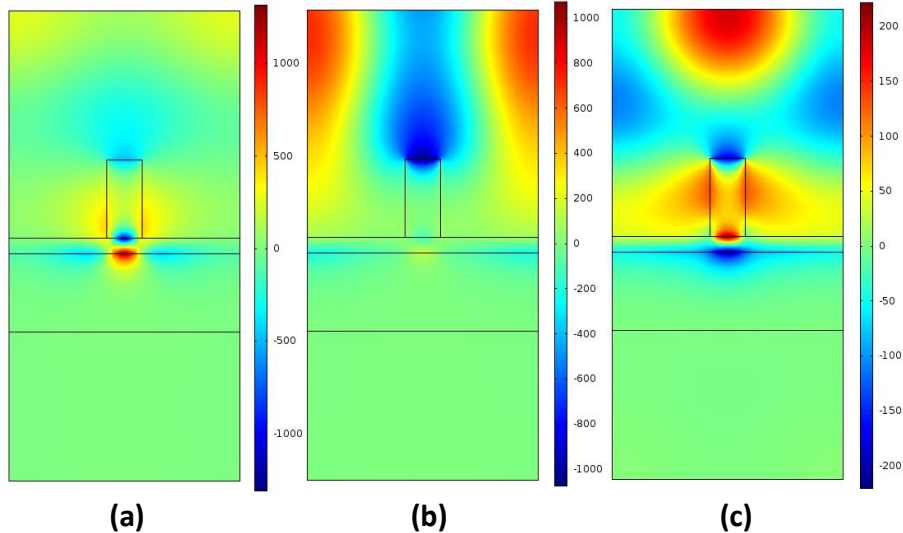

**Figure S12** $H_z$ (A/m) profile at the absorbance peaks of the strip in Fig. S10(a) for (**a**) Mode II, (**b**) Mode V, and (**c**) Mode IV.

Due to the small frequency difference between Points 4 and 5, all the resonant modes above, including near-perfect absorber mode (i.e., Mode II), can be easily blue-shifted toward Point 5 by adjusting geometric parameters. When we adjust the geometric parameters as listed in Table 1, the near-perfect semiconductor absorbance peak at Point 4 [see Fig. S11(d)] blue-shifts to Point 5 with about the same absorbance level of 95% in the semiconductor layer. The total absorbance in plain $In_{0.54}Ga_{0.46}N$ layer is between the ranges of $33\% - 35\%$ for the available experimental complex permittivity data in Fig. S2. Fig. S13 shows the effect of BSDT alone on the resonant modes, while Fig. S14 shows the effect of the lossy spacer (i.e., semiconductor). As expected, under both effects the overall behavior of the resonant modes is clearly similar to the case for Point 4 (i.e., compare Fig. S13 with Figs. S7 and S9; and Fig. S14 with Fig. S11).

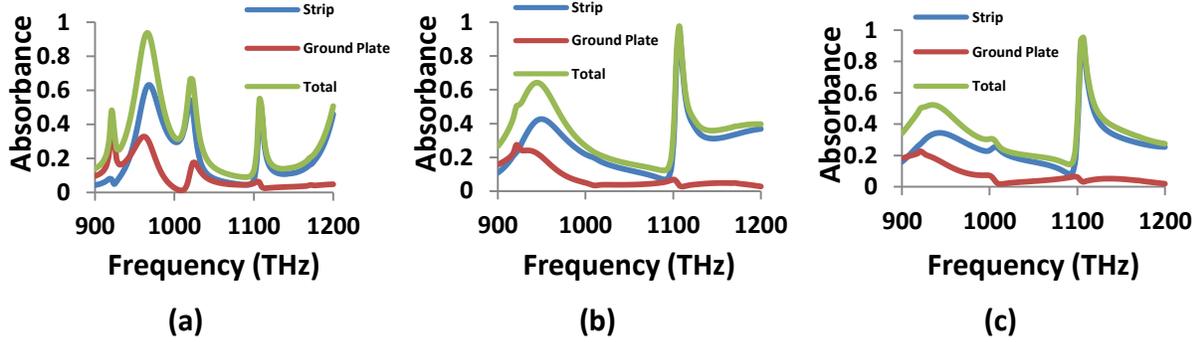

**Figure S13** (**a**) Absorbance spectra in the neighborhood of Point 5 for different parts of the absorber. The second resonance gives near-perfect absorbance. All the geometric parameters are listed in Table 1. (**b**) Absorbance spectra when (**a**) $t_s = t_g = 70$nm, (**b**) $t_s = t_g = 95$nm, and (**c**) $t_s = t_g = 105$nm. All the other parameters are kept fixed and the same as in Table 1.

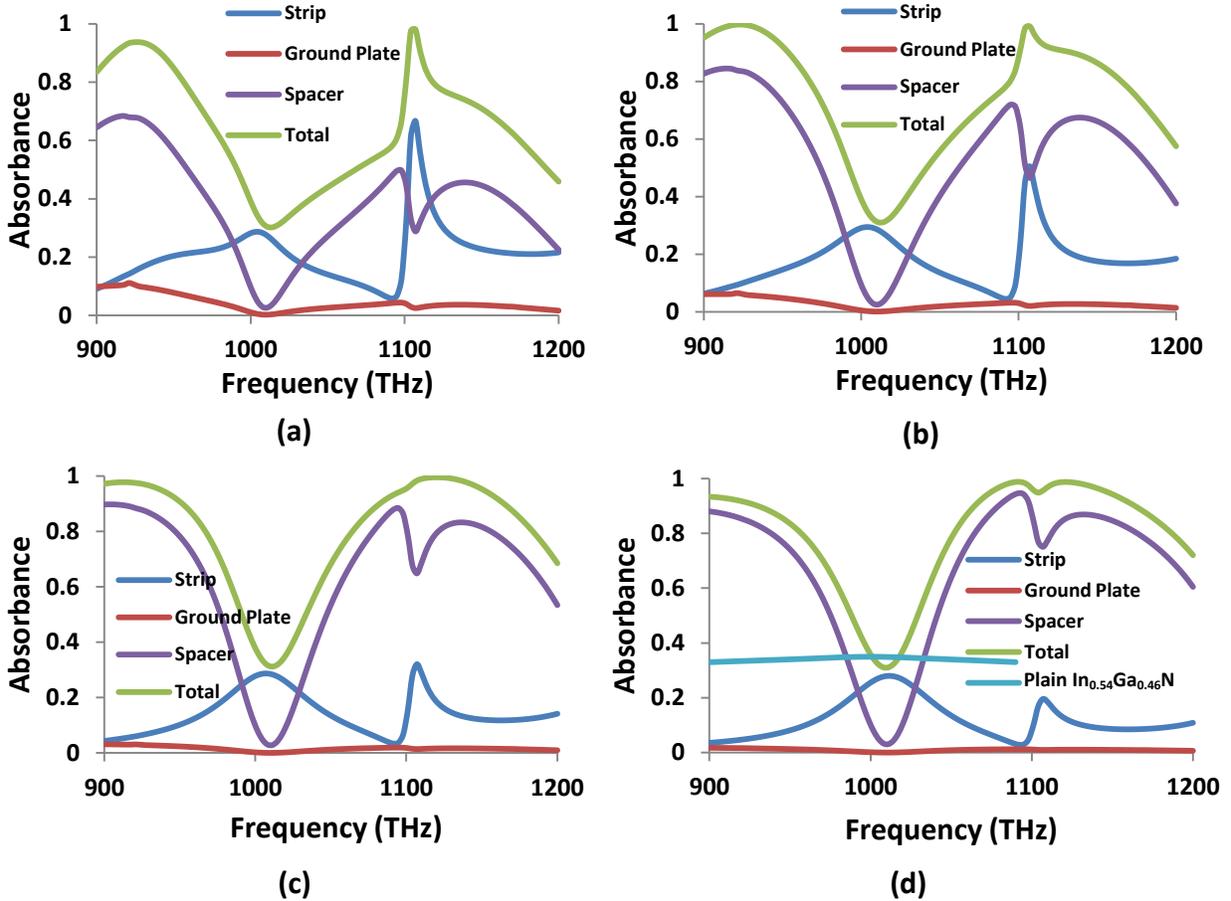

**Figure S14** Absorbance in different parts of the absorber with the parameters same as in Fig. S13(c) except for (**a**) $\Im(\varepsilon) = 0.4$, (**b**) $\Im(\varepsilon) = 0.8$, (**c**) $\Im(\varepsilon) = 1.6$, and (**d**) $\Im(\varepsilon) = \Im(\varepsilon_5) = 2.6$. The total absorbance in plain $In_{0.54}Ga_{0.46}N$ layer of the same thickness (i.e., $d = 18$nm) is also shown for comparison based on available experimental complex permittivity data in Fig. S2.